\documentclass[letterpaper,12pt]{article} 
\usepackage[]{algorithm2e}
\usepackage{amsmath}
\allowdisplaybreaks
\usepackage[usenames,dvipsnames]{color}
\usepackage[hmargin=0.7in,vmargin=1in]{geometry}
\usepackage[comma,longnamesfirst]{natbib}
\usepackage{amsfonts}
\usepackage{float}
\usepackage{amsmath, amssymb}
\usepackage{graphicx}
\usepackage{caption}
\usepackage{blindtext}
\usepackage[utf8]{inputenc}
\usepackage{setspace}
\usepackage{pdflscape}
\usepackage{numprint} 
\usepackage{setspace}

\usepackage{fontenc}
\usepackage{epsfig}
\usepackage{rotating}
\usepackage{hyperref}
\usepackage{multirow}
\usepackage{bm}

\epsfverbosetrue
\usepackage{enumitem}
\newlist{props}{enumerate}{1}
\setlist[props,1]{
	label={\arabic*.},
	leftmargin=*,
	align=left,
	labelsep=5pt,
}

\makeatletter
\def\@seccntformat#1{\@ifundefined{#1@cntformat}%
   {\csname the#1\endcsname\quad}  
   {\csname #1@cntformat\endcsname}
}
\let\oldappendix\appendix 
\renewcommand\appendix{%
    \oldappendix
    \newcommand{\section@cntformat}{\appendixname~\thesection\quad}
}
\makeatother

\usepackage{listings}
\usepackage{color} 
\definecolor{mygreen}{RGB}{28,172,0} 
\definecolor{mylilas}{RGB}{170,55,241}
\usepackage[mathlines,displaymath]{lineno}
\def \calL {\mathcal L}

\def \dvec {\text{\boldmath$d$}}

\def \uvec {\text{\boldmath$u$}}    
\def \vvec {\text{\boldmath$v$}}

\def \yvec {\text{\boldmath$y$}}    
\def \zvec {\text{\boldmath$z$}}

\def \alphavec        {\text{\boldmath$\alpha$}}

\def \gammavec        {\text{\boldmath$\gamma$}}

\def \epsilonvec      {\text{\boldmath$\epsilon$}}
\def \varepsilonvec   {\text{\boldmath$\varepsilon$}}

\def \etavec          {\text{\boldmath$\eta$}}
\def \thetavec        {\text{\boldmath$\theta$}}

\def \lambdavec       {\text{\boldmath$\lambda$}}
\def \muvec           {\text{\boldmath$\mu$}}

\def \pivec           {\text{\boldmath$\pi$}}

\def \rhovec          {\text{\boldmath$\rho$}}

\def \psivec          {\text{\boldmath$\psi$}}

\begin{document}
\pagestyle{empty}
\begin{titlepage}
\title{High-dimensional Copula Variational Approximation through Transformation}
\author{Michael Stanley Smith, Rub\'{e}n Loaiza-Maya \& David J. Nott}
\date{\today}
\maketitle
\noindent {\small Michael Stanley Smith is Chair of Management (Econometrics)
	at Melbourne Business School, University of Melbourne;
	Rub\'{e}n Loaiza-Maya is a Postdoctoral Fellow at the Department of Econometrics and Business Statistics,
Monash University;
 and,  David J. Nott is 
Associate Professor of Statistics, National University of Singapore.  
Correspondence should be directed to Michael Smith at
{\tt mike.smith@mbs.edu}.}\\
\vspace{3cm}

\noindent
{\bf Acknowledgments}: The authors would like to thank Dr. Linda Tan for providing the MCMC
output for the examples in Section~\ref{sec:skcopva}, and 
Prof. Richard Gerlach and the review team for comments
that helped improve the paper.

\newpage
\begin{center}
{\LARGE High-dimensional Copula Variational Approximation through Transformation}\\
\vspace{15pt}
{\Large Abstract}
\end{center}
\vspace{-10pt}
\noindent
Variational methods are attractive for computing Bayesian inference  when 
exact inference is impractical. 
They approximate a target distribution---either the posterior or an augmented posterior---using
a simpler distribution  that is selected to balance accuracy with computational feasibility.
Here we approximate an element-wise parametric transformation of the target distribution as multivariate Gaussian or skew-normal. 
Approximations of this kind are 
implicit copula models for the original parameters,
with a Gaussian or skew-normal copula function and flexible parametric margins. 
A key observation is that their adoption
can improve the accuracy of variational inference in high dimensions at limited or
no additional computational cost. 
We consider the Yeo-Johnson and inverse G\&H transformations, along with sparse
factor structures for the scale matrix of the Gaussian or skew-normal.
We also show how to implement efficient re-parametrization gradient methods for these 
copula-based approximations.  
The efficacy of the approach is illustrated by computing posterior inference for three different
models using
six real datasets.
In each case, we show that our proposed copula model distributions
are more accurate variational approximations
than Gaussian or skew-normal distributions, but at only a 
minor or no increase in computational cost. 
\vspace{2cm}

\vspace{1cm}

\end{titlepage}

\newpage
\pagestyle{plain}
\newpage
\doublespacing
\setlength{\abovedisplayskip}{0.15cm}
\setlength{\belowdisplayskip}{0.15cm}
\vspace{-15pt}
\section{Introduction and Literature Review}\label{sec:intro}
\vspace{-10pt}
Variational methods are an increasingly popular tool for computing posterior inferences for models with large numbers of parameters and/or large datasets; see~\cite{ormerod2010explaining} and~\cite{blei2017} for overviews.
Unlike conventional Monte Carlo methods, which are able in principle to estimate quantities of interest with any desired precision, variational methods are approximate. However, they are often substantially faster, and can be used to estimate models where exact inference is impractical.
Key to the success of variational inference is the selection of an approximation that balances accuracy with computational viability.
In this paper we suggest a general 
approach to variational inference for a high-dimensional target distribution  
using Gaussian or skew-normal copula-based approximations. They are formed
by using Gaussian or skew-normal distributions for an
element-wise parametric transformation of the target. 
Parsimonious factor parametrizations of the scale matrix of these distributions are used
to make the computations feasible.
For the transformations, we
consider the Yeo-Johnson \citep{yeojohnson2000} and inverse G\&H families~\citep{tukey77}. 
They allow for skewness and more complex features in the marginal densities of the copula model, 
without requiring a large number of additional variational parameters-- which is important for maintaining computational efficiency in high dimensions. 
We also show how efficient re-parameterization gradient methods can be
used for the copula models, including for the skew-normal by
making use of its latent Gaussian structure.
We show in a number of examples that our Gaussian and skew-normal copula models
are more accurate approximations than the corresponding Gaussian and skew-normal distributions.
Importantly, this increase in accuracy usually 
comes at only a minor increase in computational time, 
while in some instances the copula models are actually faster to calibrate. 

Variational inference methods for Bayesian computation approximate a target posterior or augmented posterior distribution using another distribution which is more tractable.  The form of the approximation is commonly derived either from an assumed factorization of the density, or the adoption of some convenient parametric family.  In the current work, we consider parametric families of approximations, for which a Gaussian is the most common choice.  Important early work on Gaussian approximations can be found in \cite{opper2009variational}, where they considered models having a Gaussian prior and factorizing likelihood, and showed that in this class of models the number of variational parameters does not proliferate with increasing dimension.  \cite{challis2013gaussian} discussed Gaussian approximations for models where the posterior could be expressed in a certain form, and show an equivalence between local variational methods and Kullback-Leibler divergence minimization
methods in their setup.  They also considered various parametrizations of the covariance matrix based on 
the Cholesky factor for the optimization.  More recent work on Gaussian approximations has focused on stochastic 
gradient methods which largely remove any restriction on the kind of models to which the methodology applies.  
Key references here are papers by \cite{Kingma+w13} and \cite{rezende+mw14} who introduced efficient variance reduction
methods for stochastic gradient estimation in the variational optimization.  These methods will be discussed further later.  
Some similar ideas were developed independently about the same time
in \cite{titsias2014doubly} and \cite{salimans2013fixed}.  The latter authors also consider methods for Gaussian approximation able to use second derivative information from the log posterior, as well as methods for forming 
non-Gaussian approximations by making use of
hierarchical structures or mixtures of Gaussians.  \cite{kucukelbir2016automatic} consider an automatic differentiation
approach to Gaussian variational approximation which considers both diagonal and dense Cholesky parametrizations of
the covariance matrix and the use of fixed marginal transformations of parameters.  Their approach is implemented in the
statistical package Stan \citep{stan}.

A key difficulty with Gaussian approximations
is the way that the number of covariance parameters increases quadratically with the number of model parameters, making
Gaussian variational approximation impractical unless more parsimonious parametrizations of the covariance matrix are adopted.  
While assuming a diagonal covariance matrix is one possibility, this leads to the inability to represent the posterior dependence.  
Work on structured approximations for covariance matrices in Gaussian approximation applicable to high-dimensional problems 
includes the work of \cite{challis2013gaussian} mentioned above, and \cite{tan2017}, who parameterize the covariance matrix in terms of a sparse Cholesky factor of the precision matrix.
Related methods for time series models are developed
in \cite{Archer2016}. \cite{Miller2016} and \cite{ong2017gaussian} 
consider factor parametrizations of covariance matrices, with the 
former authors also considering mixture approximations, with Gaussian component covariance matrices having the factor
structure.   Earlier approaches which used a one factor approximation to the covariance or precision matrix were considered by \cite{Seeger2000} and \cite{rezende+mw14}.  \cite{quiroz+nk18} consider combining factor parametrizations for state reduction with sparse precision Cholesky
factors for capturing dynamic dependence structure in high-dimensional state space models.  \cite{Guo2016} consider similar ``variational boosting'' mixture approximations to \cite{Miller2016}, although they use different approaches to the specification of mixture components and to the optimization.

The references above relate to different approaches to variational inference based on Gaussian or mixtures of Gaussians approximations.
However, there is also a large literature on other approaches to developing flexible variational families.  Most pertinent to the present work are methods based on copulas. \cite{tran+ba15} use
vine copulas, but these can be too slow to evaluate in high dimensions, and selection of the
appropriate vine structure and component pair-copulas
is difficult in general.
\cite{han2016} also employ element-wise transformations to construct
a Gaussian copula model, and their work is most closely related to ours. They consider dense Cholesky factor parametrizations for the covariance matrix in the copula, and employ approximations to 
the posterior marginals based on flexible Bernstein polynomial transformations. Our work differs from theirs in the focus on approximations that can be calibrated in high dimensions. 
In particular, we use parsimonious factor parametrizations for the copula scale matrix which are feasible to implement for a high-dimensional model parameter vector, as well as parametric transformations which are computationally efficient and do not employ too many variational parameters. We also go beyond Gaussian copula approximations by investigating skew-normal copulas as well. Skew-normal variational families are considered in \cite{ormerod11}, who considers application to models which have a structure where the lower bound can be computed using one-dimensional quadrature.  However, \cite{ormerod11} does not consider skew-normal copulas.  

Apart from copulas, there are many other ways to specify rich variational families.  These include normalizing flows \citep{rezende+m15}, Stein variational gradient descent \citep{liu2016}, real-valued non-volume preserving transformations \citep{dinh+sb16}, methods based on transport maps \citep{spantini+bm18}, implicit variational approximations where the variational family is specified through a generative process without a closed form density \citep{Huszar17} and hierarchical variational models \citep{ranganath16}.  Some of these approaches attain their flexibility through using compositions of 
transformations of an initial density, but
they do not fit into the copula framework discussed here.   

The rest of the paper is organized as follows. Section~\ref{sec:vb} gives a brief introduction to variational inference methods, followed by a general description of our proposed implicit copula approach. Sections~\ref{sec:gcopva} and~\ref{sec:skcopva} consider Gaussian copula and skew-normal copula approximations, respectively. They illustrate our approach in six examples, where the approximations are more accurate than the
corresponding Gaussian approximations, but at limited or no computational cost. 
Section~\ref{sec:discuss} gives some concluding discussion and directions for future work. MATLAB code to implement our approach is described
in the Online Appendix.

\vspace{-15pt}
\section{Variational Inference}\label{sec:vb}
\vspace{-10pt}
In this section we first provide a short overview of variational inference. We then
outline the implicit copulas formed through transformation that we employ as variational 
approximations.
\vspace{-10pt}
\subsection{Approximate Bayesian inference}
\vspace{-5pt}
We consider Bayesian inference with data $\yvec$ having density $p(\yvec|\thetavec)$, where $\thetavec=(\theta_1,\dots, \theta_m)^\top$ is either a parameter vector, or a parameter
vector augmented with some additional latent variables.  The prior
and posterior densities are denoted by $p(\thetavec)$ and $p(\thetavec|\yvec)\propto p(\thetavec)p(\yvec|\thetavec)=g(\thetavec)$, respectively.  We will consider variational inference methods, 
in which a member $q_\lambda(\thetavec)$ of some parametric family of densities is used to 
approximate $p(\thetavec|\yvec)$, where $\lambdavec\in \Lambda$ is a vector of variational parameters.
For example, for the Gaussian family
$\lambdavec$ would consist of the distinct elements of the mean vector and covariance matrix.  Approximate Bayesian inference is then formulated as an optimization
problem, where a measure of divergence between $q_\lambda(\thetavec)$ and $p(\thetavec|\yvec)$ 
is minimized with respect to $\lambdavec$. The  
Kullback-Leibler divergence
\begin{align*}
  \text{KL}(q_\lambda(\thetavec)||p(\thetavec|\yvec) ) & = \int \log \frac{q_\lambda(\thetavec)}{p(\thetavec|\yvec)} q_\lambda(\thetavec) \,d\thetavec\,,
\end{align*}
is typically used, and we employ it here.
If $p(\yvec)=\int p(\thetavec)p(\yvec|\thetavec) d\thetavec$ denotes the marginal likelihood, then it is easily shown
(see, for example, \citet{ormerod2010explaining}) that
\begin{align}
  \text{KL}(q_\lambda(\thetavec) ||p(\thetavec|\yvec) ) & = \log p(\yvec)-\int \log \frac{p(\thetavec)p(\yvec|\thetavec)}{q_\lambda(\thetavec)} q_\lambda(\thetavec) d\thetavec \nonumber \\
 & = \log p(\yvec)-\calL(\lambdavec), \label{kldexpression}
\end{align}
where $\calL(\lambdavec)$ is called the variational lower bound. Because $\log p(\yvec)$ does not depend on $\lambdavec$, minimization of the Kullback-Leibler
divergence above with respect to $\lambdavec$ is equivalent to maximizing the variational lower bound 
$\calL(\lambdavec)$. 

The lower bound takes the form of an intractable integral, so it seems challenging to optimize. However, notice
that from~\eqref{kldexpression} it can be written as an expectation with respect to $q_\lambda$ as 
\begin{equation}
\calL(\lambdavec)=E_{q_\lambda}\left[\log g(\thetavec) - \log q_\lambda(\thetavec)\right]\,,
\label{eq:lowerbound}
\end{equation}
which allows easily application of stochastic gradient ascent (SGA) methods \citep{robbins1951stochastic,bottou10}.  
In SGA we start from an initial
value $\lambdavec^{(0)}$ for $\lambdavec$ and update it recursively as
\begin{align*}
  \lambdavec^{(i+1)} & = \lambdavec^{(i)}+\rhovec_i \circ \widehat{\nabla_\lambda \calL(\lambdavec^{(i)})},
  \;\mbox{ for } i=1,2,\ldots\,,
\end{align*}
where $\rhovec_i=(\rho_{i1},\dots, \rho_{im})^\top$ is a vector of step sizes, `$\circ$' denotes the element-wise product of two vectors, and $\widehat{\nabla_\lambda \calL(\lambdavec^{(i)})}$ is an unbiased estimate of the gradient of $\calL(\lambdavec)$ at $\lambdavec=\lambdavec^{(i)}$.  For appropriate step size
choices this will converge to a local mode of $\calL(\lambdavec)$.
Adaptive step size choices are often used in practice, and we use
the ADADELTA method of \cite{zeiler2012adadelta}.  

To implement SGA unbiased estimates of the gradient of the lower bound are required.
These can be obtained directly by differentiating~\eqref{eq:lowerbound}, and evaluating the expectation
in a Monte Carlo fashion 
by simulating from $q_\lambda$.
However, variance reduction methods for
the gradient estimation are often also important for fast convergence and stability.  One of the most useful is the `reparametrization
trick' \citep{Kingma+w13,rezende+mw14}.  In this approach, 
it is assumed that an iterate $\thetavec$ can be generated from $q_\lambda$ by first drawing $\varepsilonvec$ from a density $f_\varepsilon$ which does not depend on $\lambdavec$, and then transforming $\varepsilonvec$ by a deterministic function $\thetavec=h(\varepsilonvec,\lambdavec)$ of $\varepsilonvec$ and $\lambdavec$.  From~\eqref{eq:lowerbound}, the lower bound can be written as the following
 expectation with respect to $f_\varepsilon$:
\begin{align}
 \calL(\lambdavec) & = E_{f_\varepsilon}\left[\log g(h(\varepsilonvec,\lambdavec))-\log q_\lambda(h(\varepsilonvec,\lambdavec))\right]\,. \label{lbdrepar}
\end{align}
Differentiating under the integral
sign in (\ref{lbdrepar}) gives
\begin{align}
  \nabla_\lambda \calL(\lambdavec) & = E_{f_\varepsilon}\left[\nabla_\lambda \left\{ \log g(h(\varepsilonvec,\lambdavec))-\log q_\lambda(h(\varepsilonvec,\lambdavec))\right\}\right]\,, \label{lbdgradexpr}
\end{align}
and approximating the expression (\ref{lbdgradexpr}) by Monte Carlo using one or more random draws from $f_\varepsilon$ gives an unbiased
estimate of $\nabla_\lambda \calL(\lambdavec)$. 
An intuitive reason for the success of the re-parameterization trick is that it allows gradient information from the log-posterior to be used, by moving the variational parameters
inside $g(\thetavec)$ in (\ref{lbdrepar}). \cite{xu2018} show how the trick
reduces the variance of the gradient estimates when $q_\lambda$ is a Gaussian with diagonal 
covariance matrix (the so-called 
`mean field' Gaussian approximation). We employ the 
re-parameterization trick, and specify a function $h$, for a skew-normal copula
in Section~\ref{sec:skcopva}. 


\vspace{-10pt}
\subsection{Variational approximations through transformations}\label{sec:VBwithTransformation}
\vspace{-5pt}
Let $t_{\gamma}$ be a family of one-to-one transformations onto the real line
with  
parameter vector $\gammavec$. To construct our variational approximation,
we transform each parameter as $\psi_i=t_{\gamma_i}(\theta_i)$ and 
adopt a known distribution function
$F(\psivec;\pivec)$, with
vector of parameters $\pivec$, for $\psivec=(\psi_1,\ldots,\psi_m)^\top$ . 
For example, if $F$ is a Gaussian distribution function, then $\pivec=(\muvec_\psi^\top,\mbox{vech}(\Sigma_\psi))^\top$, where $\muvec_\psi$ and $\Sigma_\psi$ are the mean and 
covariance matrix. If $p(\psivec;\pivec)=\frac{\partial^m}{\partial \psi_1\cdots\partial \psi_m}F(\psivec;\pivec)$, then 
the density of the approximation
can be recovered by computing the Jacobian of the element-wise transformation from $\thetavec$ to $\psivec$, so that
\begin{equation}
q_\lambda(\thetavec)=p(\psivec;\pivec)\prod_{i=1}^m t_{\gamma_i}'(\theta_i)\,,
\label{eq:q}
\end{equation} 
where the variational parameters are
$\lambdavec^\top=(\gammavec_1^\top,\ldots,\gammavec_m^\top,\pivec^\top)$ and
$t_{\gamma_i}'(\theta_i)=\frac{d\psi_i}{d\theta_i}$.
Moreover, if $F$ has known marginal 
distribution functions $F_i(\psi_i;\pivec_i)$ and densities $p_i(\psi_i;\pivec_i)$ for $i=1,\ldots,m$, with 
$\pivec_i\subseteq \pivec$, the marginal densities of the
 approximation are
\begin{equation}
q_{\lambda_i}(\theta_i)=p_i(\psi_i;\pivec_i)t_{\gamma_i}'(\theta_i)\,, \mbox{ for }i=1,\ldots,m\,,
\label{eq:qi}
\end{equation}
with $\lambdavec_i^\top=(\gammavec_i^\top,\pivec_i^\top)$ a sub-vector of $\lambdavec^\top$.

The density at ~(\ref{eq:q}) can also be represented using
its copula decomposition as follows. If $Q_{\lambda_i}(\theta_i)=\int_{-\infty}^{\theta_i}
q_{\lambda_i}(s)\mbox{d}s$ is the distribution function of $\theta_i$, then 
\begin{equation}
q_\lambda(\thetavec)=c(\uvec;\tilde{\pivec})\prod_{i=1}^m q_{\lambda_i}(\theta_i)\,,
\label{eq:qc}
\end{equation}
where $\uvec=(u_1,\ldots,u_m)^\top$, $u_i=Q_{\lambda_i}(\theta_i)$ and $c$ 
is an $m$-dimensional copula density with parameter vector $\tilde{\pivec}$. In much of the 
existing copula modeling literature, a parametric copula is selected for $c$. When this is combined
with pre-specified margins, this results in a flexible distributional
form for $q_\lambda$; for example, in the variational inference literature \cite{tran+ba15} use a vine copula. 
However, in this paper the copula is instead derived directly from~(\ref{eq:q})
and~(\ref{eq:qi}) by inverting Sklar's theorem, with copula density 
\[
c(\uvec;\tilde{\pivec})=\frac{p(\psivec;\pivec)}{\prod_{i=1}^m p_i(\psi_i;\pivec_i)}=
\frac{p\left((F_1^{-1}(u_1),\ldots,F_m^{-1}(u_m))^\top;\pivec\right)}{\prod_{i=1}^m p_i(F_i^{-1}(u_i);\pivec_i)}\,,
\] 
and copula function
\[
C(\uvec;\tilde{\pivec})=F\left(F_1^{-1}(u_1;\pivec_1),\ldots,F_m^{-1}(u_m;\pivec_m);\pivec\right)\,,
\]
determined by $F$. Such a copula is called 
an `inversion copula' \citep[pp.51--52]{Nelsen2006} or an `implicit copula'~\citep{mcneilbook2005}. 
In general, the copula parameters $\tilde{\pivec}$ are given by $\pivec$, but with additional 
constraints to ensure they are identifiable in the copula; see~\cite{SmithMan2016} for examples.
However, here the elements of $\pivec$ are also parameters
of the margins at~\eqref{eq:qi}, and this identifies $\pivec$ in $q_\lambda$ without any additional 
constraints.

The most popular choice for $F$
is a Gaussian distribution, resulting in the Gaussian copula~\citep{song2000}. More recently, there has been growing interest
in selecting other distributions, such as the skew-t distribution~\citep{demarta2005t,smith2012} or
those arising from state space models~\citep{SmithMan2016}. These can produce distributional
families for $q_\lambda$ that are more flexible in their dependence structures. 
Later, we will illustrate our approach with sparse Gaussian and skew-normal 
distributions for $F$, but note that other parametric distributions
can also be used.

We observe that the expression
at~\eqref{eq:q} is much easier to employ in variational inference than
that at~\eqref{eq:qc} for three reasons. First, as mentioned above, the constraints on $\pivec$ 
required to identify $\tilde{\pivec}$ do not need to be elucidated as $\pivec$ is fully identified
in~\eqref{eq:q}.
Second, evaluating~\eqref{eq:qc}
requires repeated computation of the vector $\uvec=(Q_{\lambda_1}(\theta_1),\ldots,Q_{\lambda_m}(\theta_m))^\top$ which involves $m$ numerical 
integrations, 
whereas evaluating~\eqref{eq:q} does not. Third, optimizing the lower bound with 
respect to $\tilde{\pivec}$ proves more difficult than the unconstrained $\pivec$; an observation
made previously by~\cite{han2016} for Gaussian copula variational approximation. 

\vspace{-10pt}
\subsection{Two transformations}
\vspace{-5pt}
Key to the success of our approach is the choice of an appropriate family of 
transformations $t_{\gamma}$. Because $\psi_i=t_{\gamma_i}(\theta_i)$ has distribution
function $F_i$, which is either Gaussian or skew-normal in our paper, we 
consider two choices that have proven successful in transforming data to near normality or symmetry.
The first is the single parameter transformation of
\cite{yeojohnson2000} (YJ hereafter), which extends the Box-Cox transformation to the entire real line. For $0<\gamma<2$, it is given by
\[
t_\gamma(\theta)=
\left\{\begin{array}{cl}
-\frac{(-\theta+1)^{2-\gamma}-1}{2-\gamma} &\mbox{if }\theta<0\\
\frac{(\theta+1)^\gamma -1}{\gamma} &\mbox{if }\theta\geq 0
\end{array} \right..
\]
The second is based on the two parameter (monotonic) G\&H transformation
of~\cite{tukey77}, an overview of which 
can be found in~\cite{headrick2008}.  This is used
to transform a standard Gaussian variable to another, which can
be asymmetric and heavy-tailed~\citep{peters2016}. Thus, the G\&H transformation is 
one {\em from} normality, so that we use it for $t_\gamma^{-1}$. For 
$\gamma=(g,0<h<1)$, set
\[
t_\gamma^{-1}(\psi)=
\left\{\begin{array}{cl}
	\frac{\exp(g\psi)-1}{g}\exp(h\psi^2/2) &\mbox{ if }g\neq 0\\
	\psi \exp(\frac{h\psi^2}{2}) &\mbox{ if }g=0
\end{array}\right.
\,,
\]
then $t_\gamma$ can be obtained by numerical inversion. 
We bound $h<1$ 
because it corresponds to a G\&H transformation from a standard Gaussian
to another random variable with a first moment that exists; see~\cite[Sec.5.1]{peters2016}.

For both transformations, $t_\gamma:\mathbb{R} \rightarrow \mathbb{R}$, so that if a parameter $\theta_i$ is constrained we first transform it to the real 
line; for example, with a scale or variance parameter we set $\theta_i$ to its
logarithm. 
Interestingly, when implementing SGA $t_\gamma$ is not evaluated,
but $t_\gamma^{-1}$ is repeatedly.
Table~\ref{tab:derivatives} provides these, along with expressions for  
derivatives with respect to the model and variational parameters that are required to implement the SGA algorithm. For both transformations these are all fast to compute.
\vspace{-15pt}
\section{Gaussian Copula Variational Approximation}\label{sec:gcopva}
\vspace{-10pt}
\subsection{Gaussian copula factor specification}\label{sec:factorgcopva}
\begin{figure}[H]
	\begin{center}
		\includegraphics[width=1\textwidth]{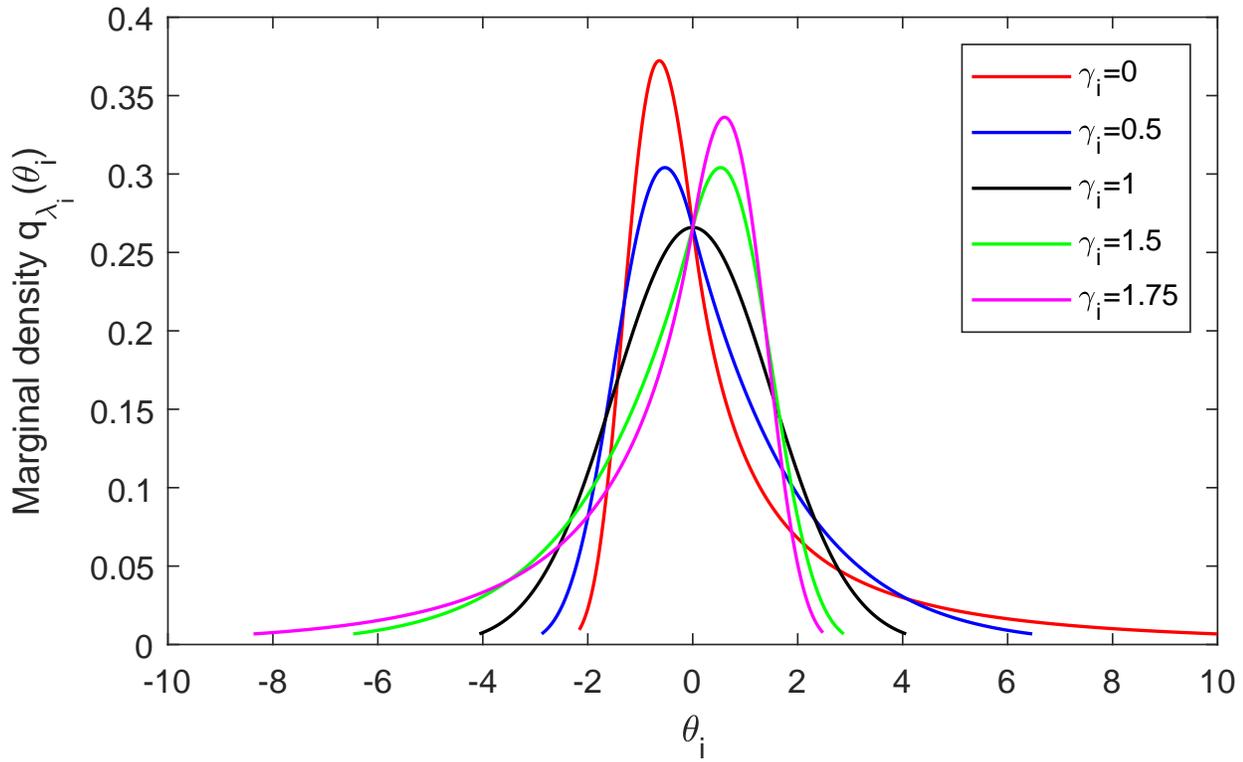}
	\end{center}
	\caption{Marginal densities $q_{\lambda_i}(\theta_i)$ of the Gaussian 
		copula variational approximation with YJ transformation. The parameters $\mu_{\psi,i}=0$ and $\sigma_{\psi,i}=1.5$, while five
		different values for $\gamma_i$ are considered.}
	\label{fig:mgcopdens}
\end{figure}

\vspace{-5pt}
The simplest implicit copula is the Gaussian copula, where $F(\psivec;\pivec)=\Phi_m(\psivec;\muvec_\psi,\Sigma_\psi)$
is a Gaussian distribution function with mean $\muvec_\psi$ and covariance matrix $\Sigma_\psi$. In constructing a Gaussian copula, it is 
usual to also set
$\muvec_\psi=(\mu_{\psi,1},\ldots,\mu_{\psi,m})^\top=\bm{0}$ and $\mbox{diag}(\Sigma_\psi)=(\sigma^2_{\psi,1},\ldots,\sigma^2_{\psi,m})=(1,1,\ldots,1)$ because these parameters
are unidentified in the Gaussian copula function;
for example, see the discussion in~\cite{song2000}. However, we do not need to do so here because
these parameters are fully identified in the density $q_\lambda$ at~\eqref{eq:q} as
they are also parameters of its margins, with
$\pivec_i=(\mu_{\psi,i},\sigma^2_{\psi,i})^\top$ at~\eqref{eq:qi}. To illustrate, Figure~\ref{fig:mgcopdens}
plots $q_{\lambda_i}$ for the YJ transformation, showing that this density can capture both 
positive or negative skew. Moreover, the direction
and level of skew can differ in each margin, depending on $\gammavec$,
making $q_\lambda$ a substantially more flexible approximation than a Gaussian.

When $\thetavec$ is of higher dimensions, we follow \cite{ong2017gaussian} and adopt a factor 
structure for $\Sigma_\psi$ as follows.  Let $B$ be an $m\times k$ matrix with $k<<m$.  For identifiability reasons
it is assumed that the upper triangle of $B$ is zero.  Let $\bm{d}=(d_1,\dots, d_m)^\top$ be a vector 
of parameters with $d_i>0$, and denote by $D$ the $m\times m$ diagonal matrix with entries $\bm{d}$.  
We assume that
\begin{align}
  \Sigma_\psi & = BB^\top+D^2, \label{factor}
\end{align}
  so that the number of parameters in $\Sigma_\psi$ grows only linearly with $m$ if $k<<m$ is kept fixed. We note that this  
  copula is equivalent to the Gaussian factor copula suggested by~\cite{Murray2013} and \cite{oh2017}
  to model data, although they do not use it as a variational approximation.
  The Gaussian
  random vector has the generative representation
 $\psivec=\muvec+B\zvec+D\epsilonvec$, where $\zvec=(z_1,\dots, z_k)^\top\sim N(0,I_k)$ and $\epsilonvec\sim N(0,I_m)$. By setting $\varepsilonvec^\top=(\zvec^\top,\epsilonvec^\top)$,
 $h(\varepsilonvec,\lambdavec)=(t_{\gamma_1}^{-1}(\psi_1),\ldots,
 t_{\gamma_m}^{-1}(\psi_m))^\top$, and $\pivec=(\muvec_\psi^\top,\mbox{vech}(B)^\top,\dvec^\top)$,
 the closed form re-parameterization gradients in a Gaussian variational approximation with factor 
 covariance structure given in \cite{ong2017gaussian} can be used.\footnote{Here the `vech'
 	operator is the half-vectorization of a rectangular matrix, defined for an $(n\times K)$ matrix $A$ with $n>K$ as $\text{vech}(A)= \left(A_{1:n,1}^\top,\dots,A_{K:n,K}^\top\right)^\top$ with $A_{k:n,k} = \left(A_{k,k},\dots,A_{n,k}\right)^\top$ for $k = 1,\dots,K$.
 	}

\vspace{-10pt}
\subsection{Application: ordinal time series copula model}\label{sec:tsapp}
\vspace{-5pt}
\subsubsection{The model and extended likelihood}\label{sec:tsmod}
To illustrate our proposed variational approximation we use it to estimate a complex model with a 
complex augmented posterior, where its greater flexibility may increase the accuracy of inference compared 
to simpler approximations.
We consider the copula time series model for an ordinal-valued random vector $\bm{Y}=(Y_1,\ldots,Y_T)^\top$
proposed by \cite{LoaizaSmith2018}. 
These authors use a $T$-dimensional parsimonious copula 
with density $c^{DV}(\vvec)$, where $\vvec=(v_1,\ldots,v_T)^\top$,
to capture serial dependence in $\bm{Y}$ (this is not to be confused with the use
of another copula for the variational approximation). The time series is assumed
to be stationary with marginal distribution function $G$, which is estimated non-parametrically in an initial step using the empirical distribution function.

The time series copula employed is a parsimonious 
drawable vine (D-vine) of Markov order $p$, as given in~\cite{Smith2015}, and 
defined as follows. Let $\{V_t\}_{t=1}^T$ be a stochastic process with
$V_t=G(Y_t)$, so that $V_t$ is marginally uniform. For $s<t$, 
denote\footnote{Note that $F_V(v_t|v_s,\ldots,v_{t-1})$ is the distribution
	function of $V_t|V_s=v_s,\ldots,V_{t-1}=v_{t-1}$ evaluated at $v_t$, 
	and $F_V(v_s|v_{s+1},\ldots,v_t)$ is the 
	distribution
	function of $V_s|V_{s+1}=v_{s+1},\ldots,V_{t}=v_t$ evaluated at $v_s$.}
$v_{t|s}=F_V(v_t|v_s,\ldots,v_{t-1})$,
$v_{s|t}=F_V(v_s|v_{s+1},\ldots,v_t)$ and $v_{t|t}=v_t$,
then the D-vine copula density is the product 
\begin{equation}
c^{DV}(\vvec;\etavec)=\prod_{t=2}^T \prod_{k=1}^{\min(t-1,p)}c^{\mbox{\tiny MIX}}_{k}(v_{t-k|t-1},v_{t|t-k+1};\etavec_{k})\,,
\label{eq:dvine}
\end{equation}
of bivariate copula densities $c^{\mbox{\tiny MIX}}_{1},\ldots,c^{\mbox{\tiny MIX}}_{p}$ called
`pair-copulas' \citep{Aas2009182},  
each with individual parameter vector $\etavec_{k}$. This D-vine copula therefore has parameter vector
$\etavec=(\etavec_1^\top,\ldots,\etavec_{p}^\top)^\top$, and is parsimonious because $|\etavec|$ does not 
increase with $T$.
To capture the heteroskedasticity that exists in most ordinal-valued time series
\cite{LoaizaSmith2018} use a five parameter mixture copula for $c^{\mbox{\tiny MIX}}_{k}$, which we also use here and is 
outlined in Part~A of the Online Appendix, leading to a total of $|\etavec|=5p$ model parameters.
Given $\vvec$, the arguments  $\{v_{t|s},v_{s|t};t=2,\ldots,T,s<t\}$ 
of the pair-copulas in ~\eqref{eq:dvine}
are computed using the recursive Algorithm 1 in~\cite{Smith2015}.

It is widely known \citep{song2000,Genest2007} that the mass function $p(\yvec|\etavec)$ of this discrete-margined copula model is computationally 
intractable, so we use the extended likelihood of 
\cite{SmithKhaled2012} instead. This employs the vector
$\bm{V}=(V_1,\ldots,V_T)^\top$, such that the joint mass function of $(\bm{Y}^\top,\bm{V}^\top)$ is
\begin{equation}
p(\yvec,\vvec|\etavec)=c^{DV}(\vvec;\etavec)\prod_{t=1}^T {\cal I}(a_t\leq v_t < b_t)\,,
\label{eq:extslike}
\end{equation}
with the indicator function ${\cal I}(X)=1$ if $X$ is true, and zero otherwise.
It is straight-forward to show that the margin in $\yvec$ of \eqref{eq:extslike} is the required
mass function $p(\yvec|\etavec)$. Evaluating the extended likelihood at~\eqref{eq:extslike} avoids the computational
burden of evaluating $p(\yvec|\etavec)$ directly.

\subsubsection{The variational approximation}
We follow~\cite{LoaizaSmith2018} and
estimate the model by setting $\thetavec=(\etavec^\top,\vvec^\top)^\top$ and
approximating the augmented posterior 
$p(\thetavec|\yvec)\propto p(\yvec,\vvec|\etavec)p(\etavec)$, which uses the extended likelihood 
and a proper uniform prior $p(\etavec)$. The target distribution
therefore has dimension $m=|\thetavec|=5p+T$. These authors use the
variational approximation
$q_\lambda(\thetavec)=q_{\lambda^a}(\etavec)q_{\lambda^b}(\vvec)$, assuming
independence between $\etavec$ and $\vvec$, and a Gaussian distribution with a factor covariance structure for $q_{\lambda^a}$.
However, because each $v_t$ is constrained to $[a_t,b_t)$, it is 
transformed to the real line as $\tilde v_t =\Phi_1^{-1}((v_t-a_t)/(b_t-a_t))$, where $\Phi_1$ is the 
distribution function of a standard Gaussian,
and independent Gaussians used as approximations for $\tilde v_1,\ldots,\tilde v_T$. 

\cite{LoaizaSmith2018} label this approximation `VA2', and we extend it as follows.
For $q_{\lambda^a}$
we use a Gaussian copula  formed through
the YJ transformation with a $k$ factor structure, so that $\lambdavec^a$ has $5p(k+3)-k(k-1)/2$ elements (the unique elements in the factor decomposition
plus the YJ transformation parameters). 
For each $\tilde v_t$ we use a normal approximation after a YJ 
transformation, so that $\lambdavec^b$ has $3T$ elements (the means and variances
of the Gaussians, plus the YJ transformation parameters). 
The full set of variational parameters are $\lambdavec=(\lambdavec^a,\lambdavec^b)^\top$.
They are calibrated using Algorithm~1 of~\cite{LoaizaSmith2018}, which employs SGA with control
variates and the analytical gradient $\nabla_{\lambda} q_\lambda$; the latter of which is given in Appendix~\ref{appA} for our copula approximation outlined here.

\subsubsection{Empirical illustration: monthly counts of attempted murder}
\begin{figure}[H]
	\begin{center}
		\includegraphics[width=0.6\textwidth]{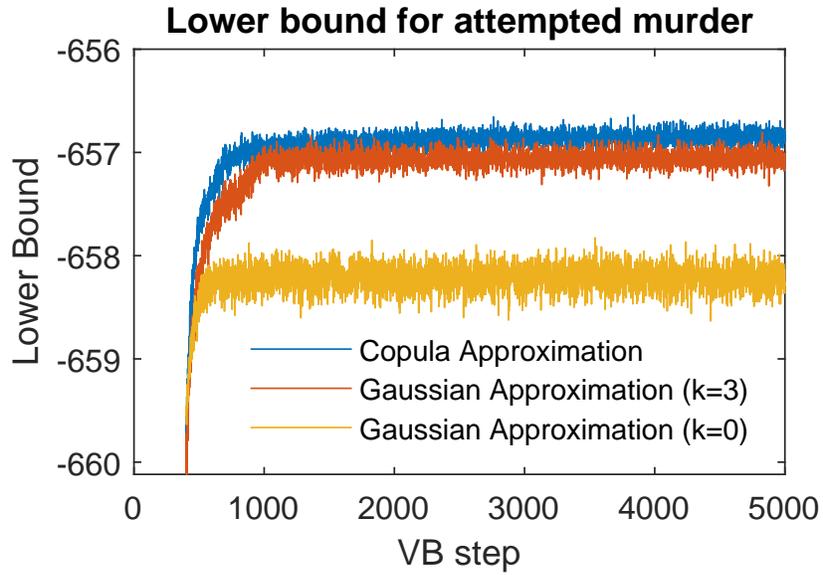}
	\end{center}
	\caption{Lower bound values for variational approximations to the posterior 
		of the copula time series model for the Attempted Murder dataset. Plot of lower bounds ${\cal L}(\bm{\lambda}^{[j]})$ against step number $j=1,\ldots,5000$ for the Gaussian copula approximation with $k=3$ factors (blue), the Gaussian approximation with $k=3$ factors (red), and the Gaussian mean field approximation (yellow).}
	\label{fig:AM1}
\end{figure}

We fit the time series model in Section~\ref{sec:tsmod} to $T=264$ monthly counts of Attempted Murder in New South Wales, Australia. Plots of the time series and 
the empirical distribution function used for margin $G$ can be found
in~\cite[Fig.1]{LoaizaSmith2018}.
The parsimonious D-vine in ~\eqref{eq:dvine}
has Markov order $p=3$, and the target density is complex with dimension $m=279$. We fit three parsimonious variational approximations: 
(i)~the Gaussian copula outlined above with $k=3$ factors, (ii)~a Gaussian distribution with factor
covariance and
$k=3$ factors,
and~(iii) a fully mean field Gaussian. Note that (ii) is equivalent to our copula
approximation but with all YJ parameters set to $\gamma_i=1$ (ie. an identity transformation), as is~(iii) but
with the additional constraint that $\Sigma_\psi$ is diagonal. 
Figure~\ref{fig:AM1} plots lower bound values against
step number for all three methods using the same SGA algorithm, and the copula approximation
clearly dominates.

\begin{figure}[H]
 	\begin{center}
 		\includegraphics[scale =0.85]{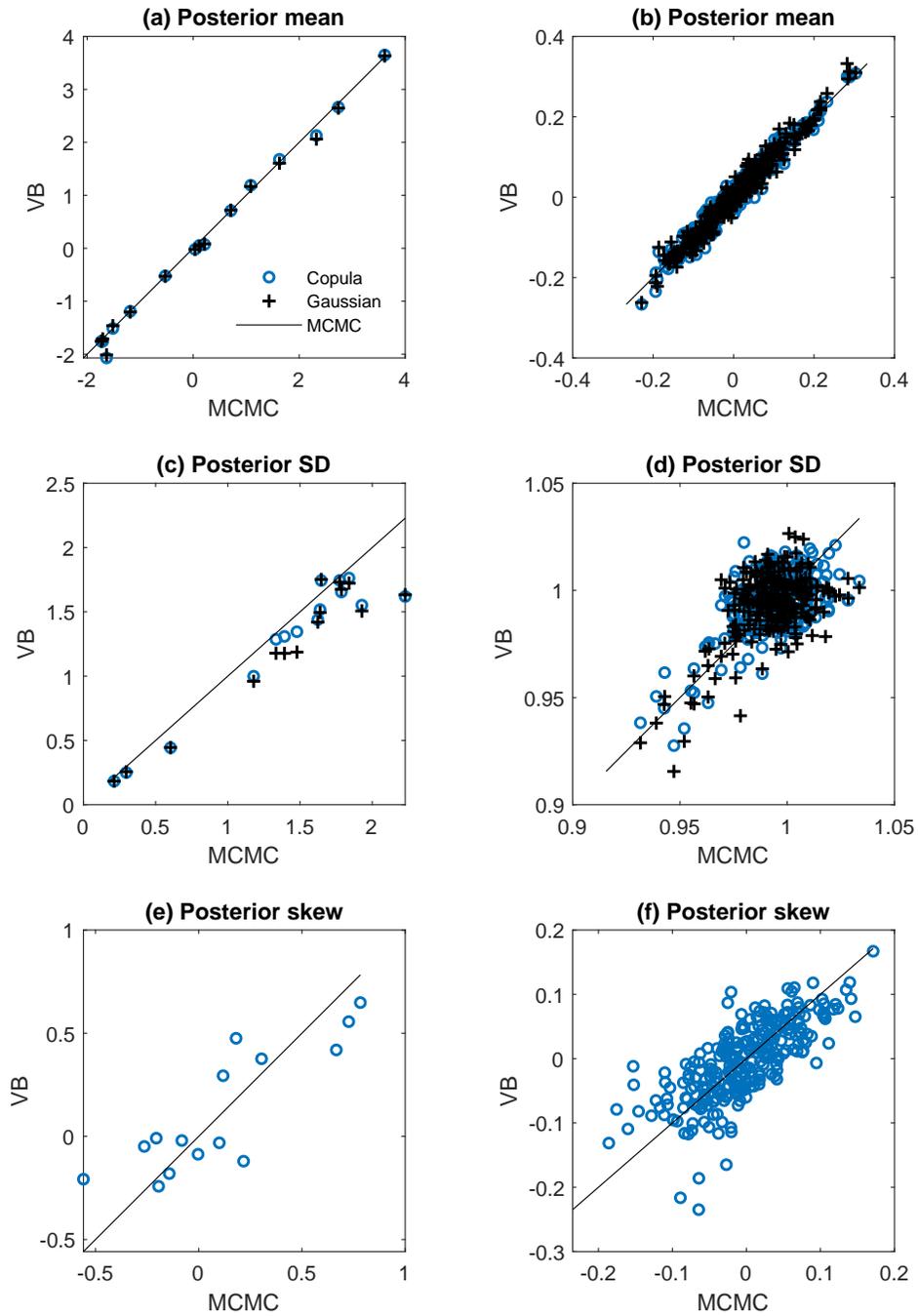}
 	\end{center}
 	\caption{Accuracy of the first three marginal posterior moments computed using
 		VB for the copula time series model fit to the Attempted Murder dataset. In each panel, the exact moment value (computed using MCMC) is plotted on the horizontal
 		axis against the moment of the variational approximation (VA) on the vertical axis. The crosses (black) are for the Gaussian VA, and the circles (blue) are for the Gaussian copula VA.  The left hand column gives the results for the (transformed) model parameters, and the right hand column gives the results for the (transformed) latent variables.}
 	\label{fig:AM2}
 \end{figure}

\begin{figure}[H]
 	\begin{center}
 		\includegraphics[scale =1]{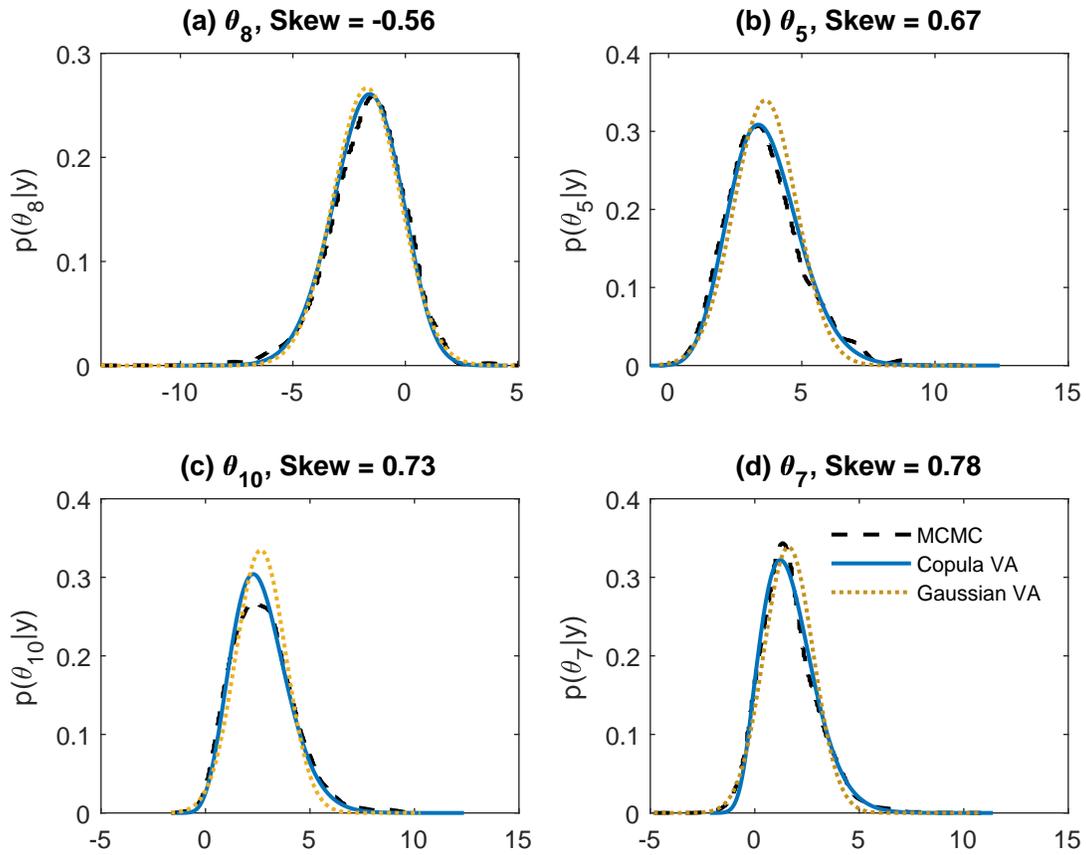}
 	\end{center}
 	\caption{Marginal posterior density estimates for the four most skewed model parameters for the copula time series model fit to the Attempted Murder dataset. In each panel the exact posterior computed using MCMC (dashed black), Gaussian copula approximation (solid blue) and Gaussian approximation (dotted yellow) are given.}
 	\label{fig:AM3}
 \end{figure}

To assess the accuracy of the three variational approximations, we also estimate the posterior
using the (slow, but exact) data augmentation 
MCMC method of~\cite{SmithKhaled2012}. Figure~\ref{fig:AM2} depicts the 
accuracy of the first three marginal posterior moments of the variational approximations. The panels provide
scatterplots of the true moments against their approximations, with a blue scatter for the proposed copula approximation,
and a red scatter for the Gaussian approximation. The left-hand
panels give results for $\etavec$ and the right-hand panel for $\vvec$.
More accurate variational approximations result in scatters
that lie closer to the 45 degree line, and we make two observations. First, panels~(e,f) show
that the true posteriors are skewed, and that the copula approximation does a very good job
of estimating the skew. Second, panel~(c) reveals that by capturing the third moment in the 
augmented vector $\thetavec=(\etavec,\vvec)$, the 
posterior standard deviation of 
$\etavec$ is also estimated more accurately. 
Figure~\ref{fig:AM3} compares the marginal densities for the four parameters
which exhibit the most skew, and the tails are more accurately estimated using the 
copula approximation.
\vspace{-15pt}
\section{Skew-Normal Copula Approximation}\label{sec:skcopva}
\vspace{-10pt}
\subsection{Copula specification}\label{sec:skcopspec}
\vspace{-5pt}
An alternative implicit copula that we consider is based on the skew-normal distribution
of~\cite{azzalini199} and \cite{azzalini2003}. In this case, the transformed parameters
$\bm\psi$ are assumed to have joint density
\begin{equation}\label{eq:skewnormdens}
p\left(\bm{\psi};\bm{\pi}\right)=2\phi_m(\bm{\psi};\bm{\mu}_\psi,\Sigma_\psi)\Phi_1(\bm\alpha_\psi^\top S_\psi^{-1/2}(\bm\psi-\bm\mu_\psi))\,,
\end{equation}
 where $\phi_m$ denotes an $m$-dimensional Gaussian density, 
 $S_\psi=\text{diag}(\sigma^2_{\psi,1},\dots,\sigma^2_{\psi,m})$, and $\sigma^2_{\psi,i}$ is the ith diagonal element of $\Sigma_\psi$. The parameters $\alphavec_\psi$ determine the level of skew
 in the marginals of $\psivec$, and when $\alphavec_\psi=\bm{0}$ the distribution reduces to a
 Gaussian. As noted in Section~\ref{sec:VBwithTransformation},
the parameters $\{\bm{\mu}_\psi,\Sigma_\psi,\bm\alpha_\psi\}$ are fully identified in the representation of $q_\lambda$ 
at~\eqref{eq:q},
whereas they are not if~\eqref{eq:skewnormdens} is used only for the construction of the copula.

\cite{demarta2005t,smith2012} and~\cite{yoshiba2018} show that implicit
copulas constructed from skew-elliptical distributions are more 
flexible than elliptical copulas because they allow for 
asymmetric dependence.\footnote{This is not to be confused with asymmetry of the marginal
distributions $q_{\lambda_i}$.}
 Here, we focus on the skew-normal copula 
because it is typically faster and easier to calibrate than the skew-t copula. When 
$\alphavec_\psi\neq \bm{0}$
it captures asymmetric dependence, making it more flexible than the Gaussian copula considered in Section~\ref{sec:gcopva}, although the same factor structure 
discussed in Section~\ref{sec:factorgcopva} is adopted for the scale matrix $\Sigma_{\psi}$.
Therefore, the approximation $q_\lambda(\thetavec)$ to the target $p(\thetavec|\yvec)$
has variational parameters
$\bm\lambda=(\bm\mu_\psi^\top,\text{vech}(B)^\top,\bm{d}^\top,\bm\alpha_\psi^\top,\bm\gamma^\top)^\top$,
where $B$ and $\dvec$ are as defined in Section~\ref{sec:factorgcopva}.

In our empirical examples, we employ the re-parametrization trick to reduce the variance of the  gradient estimate. This uses a simple generative representation of $\bm\psi$ in terms of standardized random components. Using the properties of the skew-normal distribution \citep{azzalini199}, the following generative representation for $\bm{\psi}$ can be derived (see Part~B of the Online Appendix for details). If $\Omega_\psi=S_\psi^{-1/2}\Sigma_\psi S_\psi^{-1/2}$, $\bm\delta_\psi=\left(1+\bm\alpha_\psi^\top\Omega_\psi\bm\alpha_\psi\right)^{-1/2}\Omega_\psi\bm\alpha_\psi$ and $\bm{\tilde{\delta}}_\psi=S_\psi^{1/2}\bm\delta_\psi$, then
\[
\bm{\psi}=\bm{\mu}_\psi+\bm{\tilde{\delta}}_\psi|r|+\left(I-\bm{\tilde{\delta}}_\psi\bm{\tilde{\delta}}_\psi^\top\Sigma_\psi^{-1}\right)\left(B\bm{z}+D\bm{\epsilon}\right)+\sqrt{1-\bm{\tilde{\delta}}_\psi^\top\Sigma_\psi^{-1}\bm{\tilde{\delta}}_\psi}\bm{\tilde{\delta}}_\psi \varepsilon_0\,,
\]
where $r\sim N\left(0,1\right)$,  $\varepsilon_0\sim N\left(0,1\right)$, $\bm{z}\sim N\left(\bm{0},I_k\right)$, $\bm{\epsilon}\sim N\left(\bm{0},I_m\right)$, is distributed
skew-normal with density at~\eqref{eq:skewnormdens}. Setting $\varepsilonvec^\top=(r,\varepsilon_0,\zvec^\top,\epsilonvec^\top)$ and 
$h(\varepsilonvec,\lambdavec)=(t_{\gamma_1}^{-1}(\psi_1),\ldots,
t_{\gamma_m}^{-1}(\psi_m))^\top$, the gradient at~\eqref{lbdgradexpr} can be evaluated by 
first drawing $\varepsilonvec$ from an $N(\bm{0},I)$ distribution, and 
computing the derivatives analytically; see 
Appendix~\ref{AppB} for details.

\vspace{-10pt}
\subsection{Examples}
\vspace{5pt}
To illustrate the use of a skew-normal copula as a variational approximation, we
employ it to approximate the posterior of several
logistic regressions examined previously in \cite{ong2017gaussian}. 
\subsubsection{Mixed logistic regression}
The first uses the  polypharmacy longitudinal data in~\cite{hosmer2013}, which features data on 500 subjects over 7 years. The logistic regression is specified fully in~\cite{ong2017gaussian}, and
it includes 8 fixed effects (including an intercept), plus one subject-based $N(0,\exp(2\zeta))$ random effect. The following approximations are fitted
to the augmented posterior of $\thetavec$, which comprises $\zeta$, the 8 fixed effect coefficients, and
the 500 random effect values:
\begin{itemize}
	\item[(A1)] Mean Field Gaussian: independent univariate Gaussians
	\item[(A2)] Mean Field YJ Transform: independent univariate distributions with densities at~\eqref{eq:qi},
	where $p_i(\psi_i;\pivec_i)=\phi_1(\psi_i;\mu_{\psi_i},\sigma^2_{\psi_i})$ is a Gaussian density and $t_{\gamma_i}$ is a YJ transform
	\item[(A3)] Gaussian: as in \cite{ong2017gaussian}
	\item[(A4)] Skew-normal
	\item[(A5)] Gaussian Copula: as outlined in Section~\ref{sec:factorgcopva}, with $t_{\gamma_i}$ a YJ transform
	\item[(A6)] Skew-normal Copula: as outlined in Section~\ref{sec:skcopspec}, where $t_{\gamma_i}$ is a YJ transform
	\item[(A7)] Gaussian Copula: as outlined in Section~\ref{sec:factorgcopva}, with $t_{\gamma_i}$ an inverse G\&H transform
	\item[(A8)] Skew-normal Copula: as outlined in Section~\ref{sec:skcopspec}, where $t_{\gamma_i}$ is an inverse G\&H transform
\end{itemize}

In approximations A3--A8, a factor structure with $k=5$ factors is used for the variance (A3) or 
scale matrix (A4) of the distribution, or the copula parameter matrix (A5--A8). Thus, A4 extends the approximation of~\cite{ormerod11} to include
a factor scale matrix, while A5 and A7 extend the approximation
of~\cite{han2016} to have a factor copula parameter matrix and
parametric margins constructed from the two transformations.
For each approximation
Table~\ref{tab:polypharm} lists the number of variational parameters $|\bm{\lambda}|$,
average lower bound value over the last 1000 steps of the SGA algorithm, and the
time to complete 1000 steps using MATLAB on a standard laptop. Comparing the lower bound values
for A2 and A1, 
it can be seen that allowing for asymmetry in the margins improves the approximation markedly;
although using the skew-normal A4 is not as effective. The most accurate approximations
are the Gaussian copulas A5 and A7.
The time to complete 1000 SGA steps for the copula models is almost
the same as the non-copula models (e.g. A5 and A7 are only 0.5\% and 1.5\% slower than A3) making them attractive choices.

\begin{figure}[H]
	\begin{center}
		\includegraphics[scale =0.83]{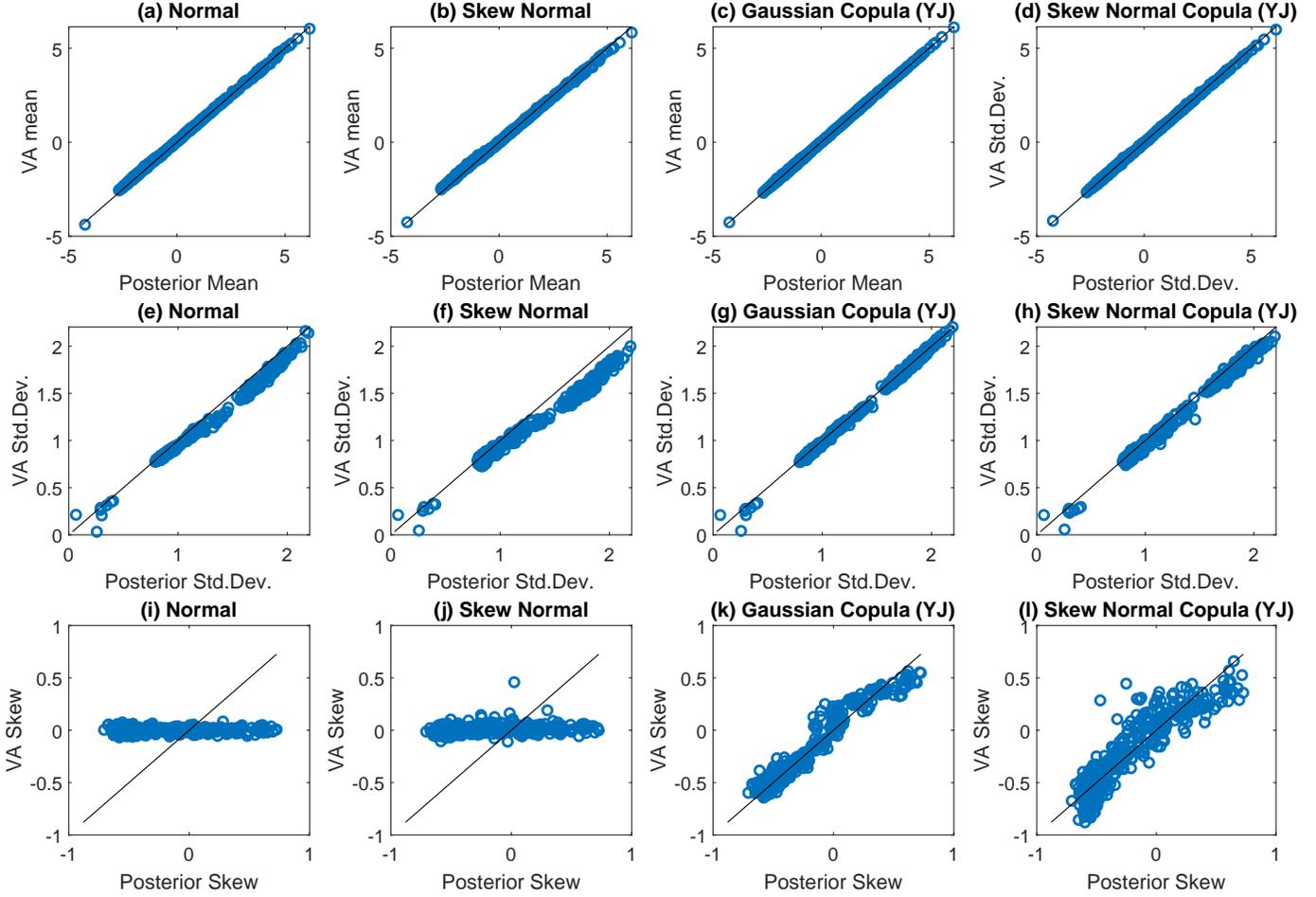}
	\end{center}
 	\caption{Accuracy of the VB estimates of the first three posterior moments of $\thetavec$
 		for the mixed logistic regression model fit to the polypharmacy dataset.
 		In each panel, the exact posterior moment (computed using MCMC) is plotted on the horizontal axis, against the equivalent moment of the VA. 
 		The means, standard deviations and 
 		Pearson's skew, are plotted in the top to bottom rows. The four columns give
 		results for four different approximations: Gaussian (A3), skew-normal (A4), Gaussian copula with YJ transform (A5), and skew-normal copula with YJ transform (A6). Each point in the scatter plot
 		correspond to an element in $\thetavec$.}
\label{fig:poly1}
\end{figure}

\begin{figure}[H]
	\begin{center}
		\includegraphics[scale =0.58]{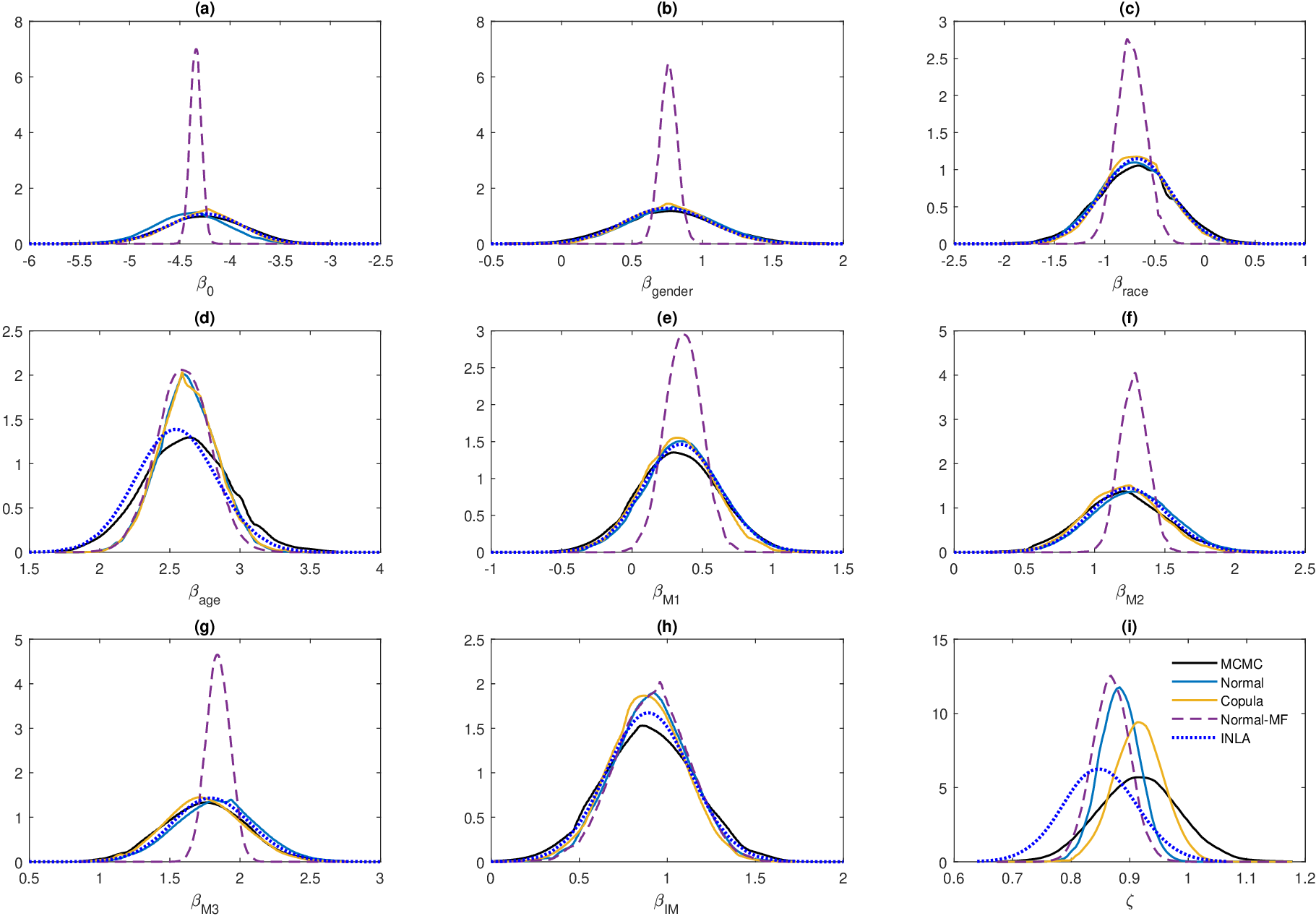}
	\end{center}
	\caption{The marginal posterior densities of the nine model parameters for
	the mixed logistic regression model fit to the polypharmacy dataset. Each panel
	plots the exact posterior computed using MCMC (black solid). The other four 
	are the approximations
	A1 mean field Gaussian (purple dashed), A3  Gaussian (blue solid), A5 Gaussian copula with YJ transform (yellow solid) and INLA (blue dotted). The densities are on the original parameter scale.}
	\label{fig:poly2}
\end{figure}

To judge the approximation accuracy, the exact augmented posterior is computed using 
MCMC with data augmentation. 
Figure~\ref{fig:poly1} plots the first three posterior moments of the approximations (vertical axes) against their true values (horizontal axes). 
Results are given for the approximations A3 (panels a,e,i), 
A4 (panels b,f,j), A5 (panels c,g,k) and A6 (panels d,h,l). 
All four identify the means well, but the striking result is that the two copula approximations 
capture the (Pearsons) skew coefficients remarkably well in panels~(k,l). 
By doing so, the estimates of the second moment in panels (g,h) are also improved. 
Figure~\ref{fig:poly2}
illustrates further by plotting the exact posterior densities for the nine model parameters (excluding the 
random effects), along with those of approximations A1, A3, A5, and that obtained using INLA~\citep{rue2009} with
the same priors. Ignoring the dependence between
parameters using A1 greatly understates the posterior standard deviation, which is well-known. However, adopting the Gaussian copula A5 improves the density estimates compared
to the Gaussian A3 -- particularly for $\zeta$ in panel~(i). The latter is likely due
to the skew in the posteriors of many random effect values, which is captured
by the copula. Last, INLA approximates the near symmetric marginal posteriors
of the fixed effects well, but has an inaccurate estimate for $\zeta$
in panel~(i), thereby understating
the level of heterogeneity in the data compared to all VB estimators.

\subsubsection{Logistic regression}
\begin{figure}[H]
 	\begin{center}
 		\includegraphics[scale=0.9]{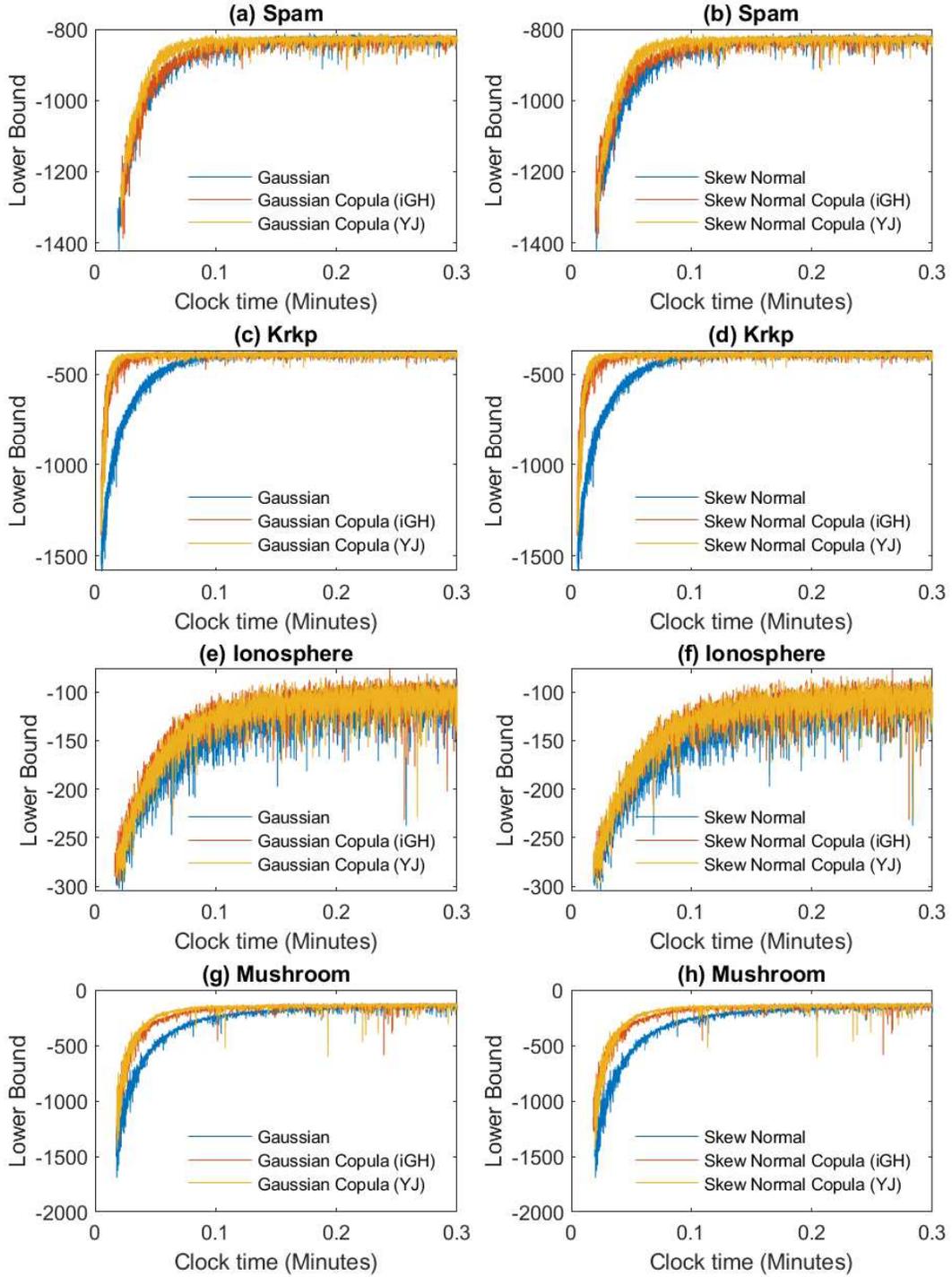}
 	\end{center}
 	\caption{Comparison of the calibration speed of different variational approximations for the four
 		logistic regression examples. Each panel plots ${\cal L}(\lambdavec)$ against the time taken to
 		 implement the SGA algorithm for three approximations. The left-hand panels give
 		plots for A3 Gaussian (blue line),  Gaussian copulas A7 (red line) and A5 (yellow line). 
 		The right-hand  panels gives plots for A4 skew-normal (blue line), skew-normal copulas A8 (red line)
 		and A6 (yellow line). For presentation purposes the results are presented only after the first 10 steps of the SGA algorithm.}
 	\label{fig:smallegs}
 \end{figure}

To illustrate the trade-off between speed and approximation accuracy, we consider the Spam, Ionosphere, Krkp and Mushroom test datasets considered in~\cite{ong2017gaussian}. These have 
sample sizes $n=4601, 351, 3196$ and $8124$, respectively, and are used to fit logistic regressions
with 104, 111, 37 and 95 covariates.
We use the same $N(0,10I)$ prior on the linear coefficients of the covariates as these authors, and fit the six
correlated approximations A3--A8 using  $k=3$ factors throughout. Table~\ref{tab:smallegs} reports the 
average lower bounds over the last 1000 steps. By this metric, 
the skewed approximations A4, A6 and A8 are the most accurate, although the differences between these three are small.  However, the copula models can have a 
substantial speed advantage. Figure~\ref{fig:smallegs} 
compares the calibration speed by plotting the lower bound against time
to implement the SGA algorithm (in MATLAB on a standard laptop). 
This shows that for the Krkp and Mushroom test
data the copula models were much faster to calibrate than either the Gaussian or skew-normal. This can also be an important consideration when using variational inference in big data problems.
\vspace{-15pt}
\section{Discussion}\label{sec:discuss}
\vspace{-10pt}
In this paper we show how to employ copula model approximations in 
variational inference using element-wise transformations of the parameter
space. This type of copula is called an `implicit copula', and is obtained from the choice of distribution $F$ for the transformed
parameters $\psivec$. We suggest using 
parametric transformations that are known to be effective
in transforming data to near normality, and illustrate with the power transformation of~\cite{yeojohnson2000}
and the inverse G\&H transformation of~\cite{tukey77}. The implied margins of such transformations
are available in closed form, and depend on both the transformation selected and the marginals 
of $F$. While, in principle, any distribution can be selected for $F$, elliptical 
and skew-elliptical \citep{genton2004} distributions are good choices for two reasons. First,
they give rise to implicit copulas which have been shown previously to be effective;
for example, see~\cite{fang2002}, \cite{demarta2005t} and \cite{smith2012}. 
Second, by employing a factor
decomposition for the scale matrix of $F$, the number of copula parameters only 
increases linearly with $m$. 

The approximation provides a balance between computational viability 
and accuracy. We illustrate here using Gaussian and skew-normal copulas 
of dimensions up to $m=509$, although higher dimensions can also be considered.
Our empirical work shows that the Yeo-Johnson transformation is particularly
effective and is quickly calibrated using SGA; in most cases, faster
than calibrating the elliptical or skew-elliptical distributions themselves on the parameter vector.
The approach of defining the copula approximation using element-wise transformations 
simplifies the computations required to implement variational inference
by using~\eqref{eq:q}. In contrast, selecting
a high-dimensional copula function---such as a vine copula~\citep{tran+ba15}---and marginals separately, uses~\eqref{eq:qc}
which is slower. 
\cite{han2016} make a similar observation
 for a Gaussian copula, and we show this applies generally to all implicit copulas. 
 Another important observation is that constraints on the parameters
of $F$ usually employed to identify the implicit copula (for example, see
~\cite{SmithMan2016}) are not required 
because they are identified through the margins $q_{\lambda_i}$.

Last, we comment on possible extensions to our work. One interesting possibility is to consider other flexible multivariate
models for constructing the implicit copula.  Truncated Gaussian graphical models \citep{su+lcc16} are one interesting possibility here, 
since they include the skew-normal distribution as a special case, and similar to the skew-normal 
they have a latent Gaussian structure which may be 
amenable to implementation of re-parametrization methods for gradient estimation in the optimization. Another interesting idea is to
use the copula Bayesian network of~\cite{elidan2010} as an approximation, where the local copulas are implicit copulas constructed
through transformation as recommended in our paper.
It would also be interesting to implement our copula approximations in other challenging settings, such as when
some of the parameters are discrete, or in likelihood-free inference applications.  Here gradient estimation
for the optimization becomes more challenging, as straightforward re-parameterization techniques do not immediately apply.

\onehalfspacing
\newpage
\appendix
\vspace{-15pt}
\section{}\label{appA}
\vspace{-10pt}
This appendix derives the gradient needed to implement the example in Section~\ref{sec:tsmod}.
In this example, $\bm{\theta}=(\bm{\eta}^\top,\bm{v}^\top)^\top$, where $\bm{\eta}$ are the
model parameter and $\bm{v}$ the vector of auxiliary variables. The approximation to the augmented posterior of $\bm{\theta}$ is
\[ q_\lambda\left(\thetavec\right)=q_{\lambda^a}\left(\bm{\eta}\right)q_{\lambda^b}\left({\bm{v}}\right)=p_a\left(\psivec^a;\bm{\pi}^a\right)p_b\left(\psivec^b;\bm{\pi}^b\right)\left(\prod_{i=1}^{m}t_{\gamma_{a,i}}'(\eta_i)\right)\left(\prod_{t=1}^{T}t_{\gamma_{b,t}}'(\tilde{v}_t)
\frac{d\tilde{v}_t}{dv_t}\right)
\]
with $\psivec^a=\left(\psi_1^a,\dots,\psi_m^a\right)^\top$, $\psi^a_i=t_{\gamma_{a,i}}(\eta_i)$,
$\psivec^b=\left(\psi_1^b,\dots,\psi_T^b\right)^\top$, $\psi_t^b=t_{\gamma_{b,t}}(\tilde{v}_t)$, 
$\tilde{v}_t=\Phi_1^{-1}\left(\frac{v_t-a_t}{b_t-a_t}\right)$,  
$\lambdavec^a=((\bm{\pi}^a)^\top,(\bm{\gamma}^a)^\top)^\top$, 
$\bm{\gamma}_a=\left(\gamma_{a,1},\dots,\gamma_{a,m}\right)^\top$, 
$\bm\lambda^b=((\bm{\pi}^b)^\top,(\bm{\gamma}^b)^\top)^\top$, 
$\bm{\gamma}_b=\left(\gamma_{b,1},\dots,\gamma_{b,T}\right)^\top$. 
It follows then that 
\[
\log q_{\lambda^a}\left(\bm{\eta}\right)=\log p_a\left(\psivec^a;\bm{\pi}_a\right)+\sum_{i=1}^m\log t_{\gamma_{a,i}}'(\eta_i)\,.
\]
For $\etavec$ we use a Gaussian copula, so that $p_a\left(\psivec^a;\bm{\pi}_a\right)=\phi_m\left(\psivec^a,\bm{\mu},BB^\top+D^2\right)$ and $\bm{\lambda}^a=\left(\bm{\mu}^\top,\bm{b}^\top,\bm{d}^\top,\bm{\gamma}^{a^\top}\right)^\top$ with $\bm b = \text{vech}(B)$ and $\bm d =\text{diag}\left(D\right)$. Following~\cite{ong2017gaussian} and~\cite{LoaizaSmith2018}, it is straightforward to 
show that the elements of the gradient 
\[\nabla_{\lambda^a} \text{log}\ q_{\lambda^a}\left(\bm{\eta}\right)=
\left(\nabla_{\mu}\text{log}\ q_{\lambda^a}\left(\bm{\eta}\right)^\top,
\nabla_{b}\text{log}\ q_{\lambda^a}\left(\bm{\eta}\right)^\top,
\nabla_{d}\text{log} \ q_{\lambda^a}\left(\bm{\eta}\right)^\top,
\nabla_{\gamma}\text{log}\ q_{\lambda^a}\left(\bm{\eta}\right)^\top\right)^\top\] are
\begin{align*}
\nabla_{\mu}\text{log} \ q_{\lambda^a}\left(\bm{\eta}\right) =& \left(BB^\top+D^2\right)^{-1}\left(\psivec^a-\bm{\mu}\right)                    \\
\nabla_{b}\text{log}  \ q_{\lambda^a}\left(\bm{\eta}\right)=&  \text{vech}\left(-\left(BB^\top+D^2\right)^{-1}B+\left(BB^\top+D^2\right)^{-1}\left(\psivec^a-\bm{\mu}\right) \left(\psivec^a-\bm{\mu}\right) ^\top\left(BB^\top+D^2\right)^{-1}B\right)                   \\
\nabla_{\bm d}\text{log} \ q_{\lambda^a}\left(\bm{\eta}\right)  =&\text{diag}\left(-\left(BB^\top+D^2\right)^{-1}D+\left(BB^\top+D^2\right)^{-1}\left(\psivec^a-\bm{\mu}\right) \left(\psivec^a-\bm{\mu}\right) ^\top\left(BB^\top+D^2\right)^{-1}D\right)\,. \\
\nabla_{\gamma}\text{log} \ q_{\lambda^a}\left(\bm{\eta}\right)=&-\frac{\partial t_\gamma\left(\bm{\eta}\right)}{\partial \bm\gamma^a} \left(BB^\top+D^2\right)^{-1}\left(\psivec^a-\bm{\mu}\right) +\left(\frac{\partial t_{\gamma_{a,1}}'(\eta_1)}{\partial\gamma_{a,1}}\frac{1}{t_{\gamma_{a,1}}'(\eta_1)},\dots,\frac{\partial t_{\gamma_{a,m}}'(\eta_m)}{\partial\gamma_{a,m}}\frac{1}{t_{\gamma_{a,m}}'(\eta_m)}\right)^\top                   
\end{align*}
with $\frac{\partial t_\gamma\left(\bm{\eta}\right)}{\partial \bm\gamma^a}=\text{Diag}\left(\frac{\partial t_{\gamma_{a,1}}(\eta_1)}{\partial\gamma_{a,1}},\dots,\frac{\partial t_{\gamma_{a,m}}(\eta_m)}{\partial\gamma_{a,m}}\right)$.\\

For $\psivec^b$  we assume an independent Gaussian approximation  $p_b\left(\psivec^b;\bm{\pi}_b\right)=\prod_{t=1}^{T}\phi_1\left(\psi^b_t;\zeta_t,\exp{\left(2c_t\right)}\right)$, where $\bm{\lambda}^b=\left(\bm{\zeta}^\top,\bm{c}^\top,\bm{\gamma}^{b^\top}\right)^\top$ , $\bm{\zeta}=\left(\zeta_1,\dots,\zeta_T\right)^\top$ and $\bm{c}=\left(c_1,\dots,c_T\right)^\top$. The implied approximation for $\bm{v}$ is
\[
\log q_{\lambda^b}(\bm{v})=\sum_{t=1}^T\left(\frac{1}{2}\tilde{v}_t^2 -c_t
-\frac{(\psi^b_t-\zeta_t)^2}{2\exp(2c_t)}-\log(b_t-a_t)+\log\left(t_{\gamma_{b,t}}'\left(\tilde{v}_t\right)\right) \right)\,,
\]
The gradient is $\nabla_{\lambda^b} \log \ q_{\lambda^b}(\bm{v})=\left(\nabla_{\zeta} \log \ q_{\lambda^b}(\bm{v})^\top,\nabla_{c} \log \ q_{\lambda^b}(\bm{v})^\top,\nabla_{\gamma} \log \ q_{\lambda^b}(\bm{v})^\top\right)^\top$ with elements            
\begin{align*}
\nabla_{\zeta} \log \ q_{\lambda^b}(\bm{v}) =&\left(\frac{\psi^b_1-\zeta_1}{\omega^2_1},\dots,\frac{\psi^b_T-\zeta_T}{\omega^2_T}\right)^\top \\
\nabla_{c} \log \ q_{\lambda^b}(\bm{v})=&\left(\frac{(\psi^b_1-\zeta_1)^2}{\omega^2_1}-1,\dots,\frac{(\psi^b_T-\zeta_T)^2}{\omega^2_T}-1\right)^\top  \\
\nabla_{\gamma} \log \ q_{\lambda^b}(\bm{v}) =&\left(\frac{1}{t_{\gamma_{b,1}}'\left(\tilde{v}_1\right)}\frac{\partial t_{\gamma_{b,1}}'\left(\tilde{v}_1\right)}{\partial\gamma_{b,1}},\dots,
\frac{1}{t_{\gamma_{b,T}}'\left(\tilde{v}_T\right)}\frac{\partial t_{\gamma_{b,T}}'\left(\tilde{v}_T\right)}{\partial\gamma_{b,T}}\right)^\top
\end{align*}            
where $\omega_t=\exp\left(c_t\right)$.    

\vspace{-15pt}
\section{}\label{AppB}
\vspace{-10pt}
This appendix provides details on the implementation of variational inference using 
the skew-normal approximation proposed in Section~\ref{sec:skcopva}. Notice that by multiplying  (\ref{eq:skewnormdens}) by
the Jacobian of the transformation from $\psivec$ to $\thetavec$, 
the approximating density is
\begin{align*}
q_\lambda(\bm\theta) & = 2\phi_m(\bm\psi;\bm\mu_\psi,\Sigma_\psi)\Phi_1(\bm\alpha_\psi^\top S_\psi^{-1/2}\left(\bm\psi-\bm\mu_\psi)\right)\prod_{i=1}^m t_{\gamma_i}'(\theta_i)\,,
\end{align*}
where $\psivec=(\psi_1,\ldots,\psi_m)^\top$ and $\psi_i=t_{\gamma_i}(\theta_i)$. The complete vector of variational parameters for this approximation is $\bm{\lambda}^\top=(\bm{\mu}_\psi^\top,\bm{\alpha}_\psi^\top,\mbox{vech}(B)^\top,\bm{d}^\top,\bm{\gamma}^\top)$, where $\mbox{vech}(B)$ is the vectorization of $B$ omitting the zero upper triangular elements.
As discussed in Section~\ref{sec:vb}, to implement SGA using the re-parameterization trick,
the gradient
\begin{align}
\nabla_\lambda \calL(\lambdavec) =& E_{f_\varepsilon}\left[\nabla_\lambda \left(\log g(h(\varepsilonvec,\lambdavec))-\log q_\lambda(h(\varepsilonvec,\lambdavec))\right)\right]\nonumber\\
=&E_{f_\varepsilon}\left[\left\{\frac{d  h(\bm\varepsilon,\bm\lambda)}{d\bm\lambda}\right\}^T\left(\nabla_\theta \log g(h(\bm\varepsilon,\bm\lambda))-\nabla_\theta \log q_\lambda(h(\bm\varepsilon,\bm\lambda))\right)\right]\,,  \label{eq:appA}
\end{align}
needs approximating. This is undertaken by drawing an iterate of $\varepsilonvec=(r,\varepsilon_0,\zvec^\top,\epsilonvec^\top)^\top$ from 
a $N(\bm{0},I)$ distribution, and then computing the derivatives inside~\eqref{eq:appA} 
analytically. Below, we write $\thetavec=h(\varepsilonvec,\lambdavec)$ as $\bm\theta(\bm{\varepsilon},\bm\lambda)$ for clarity. To derive the derivatives, note that 
the gradient can be broken up into sub-vectors
\[
\nabla_\lambda \calL(\lambdavec)=\left(\nabla_{\mu_\psi}  \calL(\lambdavec)^\top,\nabla_{\alpha_\psi}  \calL(\lambdavec)^\top,\nabla_{\mbox{\footnotesize vech}(B)}  \calL(\lambdavec)^\top,\nabla_d  \calL(\lambdavec)^\top,\nabla_\gamma \calL(\lambdavec)^\top\right)^\top\,, 
\]
where
\begin{align*}
	\nabla_{\mu_\psi}  \calL(\lambdavec) =    & \frac{d \bm\theta(\bm{\varepsilon},\bm\lambda)}{d \bm\mu_\psi}^\top\left(\nabla_\theta\log g\left(\bm\theta\right)-\nabla_\theta \log q_\lambda\left(\bm\theta\right)\right)    \\
	\nabla_{\alpha_\psi}  \calL(\lambdavec) = & \frac{d \bm\theta(\bm{\varepsilon},\bm\lambda)}{d \bm\alpha_\psi}^\top\left(\nabla_\theta\log g\left(\bm\theta\right)-\nabla_\theta \log q_\lambda\left(\bm\theta\right)\right)  \\
	\nabla_{\mbox{\footnotesize vech}(B)}  \calL(\lambdavec) =             & \frac{d \bm\theta(\bm{\varepsilon},\bm\lambda)}{d {B}}^\top\left(\nabla_\theta\log g\left(\bm\theta\right)-\nabla_\theta \log q_\lambda\left(\bm\theta\right)\right)               \\
	\nabla_d  \calL(\lambdavec) =        & \frac{d \bm\theta(\bm{\varepsilon},\bm\lambda)}{d \bm{d}}^\top\left(\nabla_\theta\log g\left(\bm\theta\right)-\nabla_\theta \log q_\lambda\left(\bm\theta\right)\right)       \\
	\nabla_\gamma \calL(\lambdavec) =         & \frac{d \bm\theta(\bm{\varepsilon},\bm\lambda)}{d \bm\gamma}^\top\left(\nabla_\theta\log g\left(\bm\theta\right)-\nabla_\theta \log q_\lambda\left(\bm\theta\right)\right) 
\end{align*}
the
derivative with respect to $\mbox{vech}(B)$ above 
is computed by ignoring elements on right hand side of the equation that
correspond to the upper triangle of $B$. The term $\nabla_\theta \log g(\bm\theta)$ is model specific and needs to be derived on a 
case-by-case basis. 
Expressions for the remaining terms can be computed in closed form.
First, 
\begin{eqnarray*}
\frac{d \bm\theta(\bm{\varepsilon},\bm\lambda)}{d \bm\mu_\psi} &= &\frac{d t_\gamma^{-1} (\bm\psi)}{d \bm\psi}=\text{diag}\left(\frac{d t_{\gamma_1}^{-1} (\psi_1)}{d \psi_1},\dots,\frac{d t_{\gamma_m}^{-1} (\psi_m)}{d \psi_m}\right)\\
\frac{d \bm\theta(\bm{\varepsilon},\bm\lambda)}{d \bm\gamma} &= &\frac{d t_\gamma^{-1} (\bm\psi)}{d \bm\gamma}=\text{diag}\left(\frac{d t_{\gamma_1}^{-1} (\psi_1)}{d \gamma_1},\dots,\frac{d t_{\gamma_m}^{-1} (\psi_m)}{d \gamma_m}\right)\,,
\end{eqnarray*}
where the elements are computed using the formulas given in Table~\ref{tab:derivatives} for either
the YJ or G\&H transformations. 
Expressions for the remaining four derivatives are provided in Table~\ref{Tab:SkewNormDerivatives}, 
which are derived in the Online Appendix. MATLAB routines that evaluate
these derivatives are in the Supplementary Material.

\vspace{-15pt}
\section*{Supplementary Materials}
\vspace{-10pt}
Supplementary materials contain:
\begin{description}
	\item[smith\_loaiza\_maya\_nott\_webappend.pdf] An online appendix in two 
parts. Part~A specifies the pair-copula used in Section~\ref{sec:tsapp}; Part~B derives the four derivatives in Appendix B.
\end{description}
\baselineskip=15pt 
\bibliography{Project_bib}
\bibliographystyle{apa}
\newpage
\begin{landscape}
	\begin{table}[H]
		\centering
		{\small	
			\resizebox{1.3\textwidth}{!}{
				\begin{tabular}{lcccc}
					\hline\hline
					                                                                                 &                                                                                                   \multicolumn{2}{c}{Yeo-Johnson Transformation}                                                                                                    &                                                                                                                                                                                          \multicolumn{2}{c}{Inverse G\&H Transformation}                                                                                                                                                                                           \\[0.1cm]
					Function                                                                         &                                                          $\theta<0,\psi<0$                                                           &                                           $\theta\ge 0,\psi\ge 0$                                            &                                                                                                      $g\ne0$                                                                                                       &                                                                                                      $g=0$                                                                                                      \\[0.1cm] \hline
					                                                                                 &                                                                                                                                      &                                                                                                              &                                                                                                                                                                                                                  &                                                                                                                                                                                                                 \\[0.05cm]
					$t_\gamma\left(\theta\right)$                                                    &                                            $-\frac{\bar{\theta}^{2-\gamma}-1}{2-\gamma}$                                             &                                    $\frac{(\theta+1)^\gamma -1}{\gamma} $                                    &                                                                                                 Evaluated Numerically                                                                                                 &                                                                                                Evaluated Numerically                                                                                             \\[0.4cm]
					$t_\gamma^{-1}\left(\psi\right)$                                                 &                                  $1-\left(1-\psi\left(2-\gamma\right)\right)^{\frac{1}{2-\gamma}}$                                   &                               $\left(1+\psi\gamma\right)^{\frac{1}{\gamma}}-1$                               &                                                                                     $\frac{\exp(g\psi)-1}{g}\exp(h\psi^2/2)$                                                                                     &                                                                                         $\psi \exp(\frac{h\psi^2}{2}) $                                                                                         \\[0.4cm]
					$\frac{\partial }{\partial\psi}t_\gamma^{-1}(\psi) $                             &                                $\left(1-\psi\left(2-\gamma\right)\right)^{\frac{\gamma-1}{2-\gamma}}$                                &                            $\left(1+\psi\gamma\right)^{\frac{1-\gamma}{\gamma}}$                             &                                                                       $\exp\left(g\psi+\frac{h\psi^2}{2}\right)+h\psi t^{-1}_\gamma(\psi)$                                                                       &                                                                         $\exp\left(\frac{h\psi^2}{2}\right)+h\psi t^{-1}_\gamma(\psi)$                                                                          \\[0.4cm]
					$t_\gamma'\left(\theta\right)=\frac{\partial }{\partial\theta}t_\gamma(\theta) $ &                                                      $\bar{\theta}^{1-\gamma}$                                                       &                                      $\left(\theta+1\right)^{\gamma-1}$                                      &                                                                      $\left[\frac{\partial }{\partial\psi}t_\gamma^{-1}(\psi)\right]^{-1}$                                                                       &                                                                      $\left[\frac{\partial }{\partial\psi}t_\gamma^{-1}(\psi)\right]^{-1}$                                                                      \\[0.4cm]
					$\frac{\partial^2 }{\partial\psi^2}t_\gamma^{-1}(\psi) $                         &                                                             Not Required                                                             &                                                 Not Required                                                 &                                                                         $\exp\left(g\psi+\frac{h\psi^2}{2}\right)\left(g+h\psi\right)+$                                                                          &                                                                                   $\exp\left(\frac{h\psi^2}{2}\right)h\psi+$                                                                                    \\[0.4cm]
					                                                                                 &                                                                                                                                      &                                                                                                              &                                                      $h\psi \left[\frac{\partial }{\partial\psi}t_\gamma^{-1}(\psi)\right]+ht_\gamma^{-1}\left(\psi\right)$                                                      &                                                     $h\psi \left[\frac{\partial }{\partial\psi}t_\gamma^{-1}(\psi)\right]+ht_\gamma^{-1}\left(\psi\right)$                                                      \\[0.4cm]
					$\frac{\partial^2 }{\partial\theta^2}t_\gamma(\theta) $                          &                                            $\left(\gamma-1\right)\bar{\theta}^{-\gamma}$                                             &                           $\left(\gamma-1\right)\left(\theta+1\right)^{\gamma-2}$                            &                                           $-\left[\frac{\partial }{\partial\psi}t_\gamma^{-1}(\psi)\right]^{-3}\frac{\partial^2 }{\partial\psi^2}t_\gamma^{-1}(\psi) $                                           &                                          $-\left[\frac{\partial }{\partial\psi}t_\gamma^{-1}(\psi)\right]^{-3}\frac{\partial^2 }{\partial\psi^2}t_\gamma^{-1}(\psi) $                                           \\[0.4cm]
					$\frac{\partial }{\partial\gamma}t_\gamma(\theta) $                              & $\frac{\left(2-\gamma\right)\bar{\theta}^{2-\gamma}\ln\left(\bar{\theta}\right)-\bar{\theta}^{2-\gamma}+1}{\left(2-\gamma\right)^2}$ & $\frac{\gamma\left(1+\theta\right)^\gamma\ln\left(\theta+1\right)-\left(1+\theta\right)^\gamma+1}{\gamma^2}$ &                                                                                                   Not Required                                                                                                   &                                                                                                  Not Required                                                                                                   \\[0.4cm]
					$\frac{\partial }{\partial\gamma}t_\gamma^{-1}(\psi) $                           &                $-\frac{\partial }{\partial\psi}t_\gamma^{-1}(\psi)\frac{\partial }{\partial \gamma}t_\gamma(\theta)$                 &    $-\frac{\partial }{\partial\psi}t_\gamma^{-1}(\psi)\frac{\partial }{\partial \gamma}t_\gamma(\theta)$     &                                      $\frac{\partial }{\partial g}t_\gamma^{-1}(\psi)=\frac{\psi}{g}\exp\left(g\psi+\frac{h\psi^2}{2}\right)-\frac{t_\gamma^{-1}(\psi)}{g}$                                      &    $\frac{\partial }{\partial h}t_\gamma^{-1}(\psi)=\frac{\psi^2}{2}t_\gamma^{-1}\left(\psi\right)$                                                                   
					\\[0.4cm]
					                                                                                 &                                                                                                                                      &                                                                                                              &                                                         $\frac{\partial }{\partial h}t_\gamma^{-1}(\psi)=\frac{\psi^2}{2}t_\gamma^{-1}\left(\psi\right)$                                                         &                                                                                                               \\[0.4cm]
					$\frac{\partial}{\partial\gamma}t_\gamma'\left(\theta\right)$                    &                                 $-\left(\bar{\theta}\right)^{1-\gamma}\ln\left(\bar{\theta}\right)$                                  &                          $\left(\theta+1\right)^{\gamma-1}\ln\left(\theta+1\right)$                          & $\frac{\partial }{\partial g}t_\gamma'(\theta) = -\left[\frac{\partial }{\partial\psi}t_\gamma^{-1}(\psi)\right]^{-2}\frac{\partial }{\partial g}\left[\frac{\partial }{\partial\psi}t_\gamma^{-1}(\psi)\right]$ &                                                                         
					\\[0.4cm]
					                                                                                 &                                                                                                                                      &                                                                                                              & $\frac{\partial }{\partial h}t_\gamma'(\theta) = -\left[\frac{\partial }{\partial\psi}t_\gamma^{-1}(\psi)\right]^{-2}\frac{\partial }{\partial h}\left[\frac{\partial }{\partial\psi}t_\gamma^{-1}(\psi)\right]$ & $\frac{\partial }{\partial h}t_\gamma'(\theta) =-\left[\frac{\partial }{\partial\psi}t_\gamma^{-1}(\psi)\right]^{-2}\frac{\partial }{\partial h}\left[\frac{\partial }{\partial\psi}t_\gamma^{-1}(\psi)\right]$\\
					                                                                          \hline\hline
				\end{tabular}
			}
		}
		\caption{Two transformations, their inverses and derivatives that are required to implement
			the copula variational Bayes estimator. For the YJ transformation, the term $\bar{\theta}=1-\theta$, and $\gamma$ is a scalar. 
			The inverse G\&H is a two parameter transformation with $\gamma=\{g,0<h<1\}$. Note that $t_\gamma$ in the first row is never computed
		in the SGA algorithm, along with a number of derivatives labelled `Not Required'. MATLAB routines to evaluate the functions are provided in the Supplementary Materials.}
		\label{tab:derivatives}
	\end{table} 
\end{landscape}

\begin{table}[H]
\centering
	{\small	
		\begin{tabular}{lccc}
			\hline\hline
			Variational Approximation               & \# Parameters $|\lambdavec|$ & Max. Lower Bound & Time (mins) \\ \hline
			(A1) Mean Field Gaussian                &             1018             &     -923.08      &    0.85     \\
			(A2) Mean Field YJ Transform            &             1527             &     -913.17      &    0.85     \\
			(A3) Gaussian                           &             3553             &     -918.24      &    1.99     \\
			(A4) Skew-normal                        &             4062             &     -923.16      &    2.28     \\
			(A5) Gaussian Copula (YJ Transform)     &             4062             &     -908.33      &    2.00     \\
			(A6) Skew-normal Copula (YJ Transform)  &             4571             &     -916.80      &    2.34     \\
			(A7) Gaussian Copula (iGH Transform)    &             4571             &     -909.21      &    1.86     \\
			(A8) Skew-normal Copula (iGH Transform) &             5080             &     -924.01      &    2.15     \\ \hline\hline
		\end{tabular}
}
	\caption{Comparison of different variational approximations
	$q_\lambda(\thetavec)$ to the augmented
	posterior of the mixed logistic regression for the polypharmacy data. The mean field
	Gaussian, with and without YJ transformation, are included as benchmarks A1 and A2.
	All the remaining approximations use factor decompositions for the scale matrices with $k=5$ factors. 
For each approximation, the number of variational parameters $|\lambdavec|$, 
average lower bound value over the last 1000 steps, and the time to complete 1,000 steps using MATLAB on a standard laptop are reported.}
\label{tab:polypharm}
\end{table}

\begin{table}[H]
	\centering
		\begin{tabular}{lcccccc}
			\hline\hline
			Example &       \multicolumn{6}{c}{Variational Approximation}       \\
			        &  (A3)   &  (A4)   &  (A5)   &  (A6)   &  (A7)   &  (A8)   \\ \hline
			Spam    & -828.15 & -824.28 & -827.96 & -824.02 & -828.69 & -824.60 \\
			Krkp    & -386.39 & -386.68 & -386.86 & -384.98 & -390.33 & -386.80 \\
			Iono    & -103.98 & -100.39 & -104.39 & -100.95 & -106.26 & -102.46 \\
			Mush    & -126.31 & -124.06 & -127.93 & -124.21 & -132.15 & -129.15 \\ \hline\hline
		\end{tabular}
	\caption{Average lower bound value over the last 1000 steps for six variational approximations to the 
		posteriors of the four logistic regression examples.}
	\label{tab:smallegs}
\end{table} 

	\begin{table}[H]
	\centering
	{\small	
		\begin{tabular}{llll}
			\hline\hline
			\multicolumn{2}{l}{Computing $\frac{d \bm\theta(\bm{\varepsilon},\bm\lambda)}{d B}$}                                                                                                               & \multicolumn{2}{l}{Computing $\frac{d \bm\theta(\bm{\varepsilon},\bm\lambda)}{d \bm{d}}$}                                                                                               \\ \hline
			&                                                                                                                          &                                             &                                                                                                                          \\
			$M_1=$                                                 & $\bm{\tilde{\delta}}_\psi\bm{\tilde{\delta}}_\psi^\top$                                                                     & $M_1=$                                      & $\bm{\tilde{\delta}}_\psi\bm{\tilde{\delta}}_\psi^\top$                                                                     \\
			$M_2=$                                                 & $\bm{\xi}^\top\Sigma_\psi^{-1}\bm{\tilde{\delta}}_\psi I_m+\bm{\xi}^\top\Sigma_\psi^{-1}\otimes \bm{\tilde{\delta}}_\psi$        & $M_2=$                                      & $\bm{\xi}^\top\Sigma_\psi^{-1}\bm{\tilde{\delta}}_\psi I_m+\bm{\xi}\top\Sigma_\psi^{-1}\otimes \bm{\tilde{\delta}}_\psi$        \\
			$M_3=$                                                 & $0.5\varepsilon_0(1-\bm{\tilde{\delta}}_\psi^\top\Sigma_\psi^{-1} \bm{\tilde{\delta}}_\psi )^{-1/2} \bm{\tilde{\delta}}_\psi$ & $M_3=$                                      & $0.5\varepsilon_0(1-\bm{\tilde{\delta}}_\psi^\top\Sigma_\psi^{-1} \bm{\tilde{\delta}}_\psi )^{-1/2} \bm{\tilde{\delta}}_\psi$ \\
			$M_4=$                                                 & $\varepsilon_0 \sqrt{1-\bm{\tilde{\delta}}_\psi^\top \Sigma_\psi^{-1}\bm{\tilde{\delta}}_\psi}$                               & $M_4=$                                      & $\varepsilon_0 \sqrt{1-\bm{\tilde{\delta}}_\psi^\top \Sigma_\psi^{-1}\bm{\tilde{\delta}}_\psi}$                               \\
			$M_5=$                                                 & $\bm{\tilde{\delta}}_\psi^\top \left(\Sigma_\psi^{-1}B\otimes \bm{\tilde{\delta}}_\psi^\top\Sigma_\psi^{-1} \right)$           & $M_{5}=$                                  & $-\left(\Sigma_\psi^{-1}D\otimes\bm{\tilde{\delta}}_\psi^\top\Sigma_\psi^{-1}+\right.$                                       \\
			$M_6=$                                                 & $\bm{\tilde{\delta}}_\psi^\top \left(\Sigma_\psi^{-1}\otimes \bm{\tilde{\delta}}_\psi^\top\Sigma_\psi^{-1} B\right)K_{m,p}$    &                                             & $\left.\bm{\tilde{\delta}}_\psi^\top\Sigma_\psi^{-1}D\otimes\Sigma_\psi^{-1}\right)P$                                        \\
			$M_{7} =$                                              & $[\text{diag}(B_{.1}),\dots ,\text{diag}(B_{.p})]$                                                                       & $M_6 =$                                     & $[\text{diag}(D_{.1}),\dots ,\text{diag}(D_{.p})]$                                                                       \\
			$M_{8} =$                                              & $\text{diag}(\bm{\delta_\psi})S_\psi^{-1/2}M_{7}$                                                                        & $M_7 =$                                     & $\text{diag}(\bm{\delta_\psi})S_\psi^{-1/2}M_6P$                                                                         \\
			$M_{9}=$                                               & $-2M_3\otimes\left(\bm{\tilde{\delta}}_\psi^\top\Sigma_\psi^{-1} M_{8}\right)$                                              & $M_8=$                                      & $-2M_3\otimes\left(\bm{\tilde{\delta}}_\psi^\top\Sigma_\psi^{-1} M_7\right)$                                                \\
			$M_{10}=$                                              & $M_3 \otimes M_5+ M_3 \otimes M_6$                                                                                       & $M_9=$                                      & $-M_3 \otimes\left(\bm{\tilde{\delta}}_\psi^\top M_{5}\right) $                                                              \\
			$M_{11} =$                                             & $M_2M_{8}$                                                                                                               & $M_{10} =$                                  & $M_2M_7$                                                                                                                 \\
			$M_{12} =$                                             & $ -(\bm{\xi}^\top\Sigma_\psi^{-1}B \otimes \Sigma_\psi^{-1})-$                                                               & $M_{11} =$                                & $ -\left((\bm{\xi}^\top\Sigma_\psi^{-1}D \otimes \Sigma_\psi^{-1})+\right.$                                                  \\
			& $K_{1,m}(\Sigma_\psi^{-1}B\otimes\bm{\xi}^\top\Sigma_\psi^{-1} )$                                                            &                                             & $\left.K_{1,m}(\Sigma_\psi^{-1}D\otimes\bm{\xi}^\top\Sigma_\psi^{-1} )\right)P$                                              \\
			$T_{B0} =$                                             & $|r|M_8$                                                                                                                 & $T_{d 0} =$                            & $|r|M_7 $                                                                                                                \\
			$T_{B1} =$                                             & $\bm{z}^\top\otimes I_m$                                                                                                    & $T_{d 1} =$                            & $\left(\bm{\epsilon}^\top\otimes I_m\right)P$                                                                               \\
			$M_{13} =$                                             & $M_{12}+\Sigma^{-1}_\psi T_{B1} $                                                                                        & $M_{12} =$                                  & $M_{11}+\Sigma^{-1}_\psi T_{d 1} $                                                                                  \\
			$T_{B2}=$                                              & $M_{11}+M_1M_{13}$                                                                                                       & $T_{d 2}=$                             & $M_{10}+M_1M_{12}$                                                                                                       \\
			$T_{B3}=$                                              & $ M_{9}+M_{10}+ M_4 M_{8}$                                                                                               & $T_{d 3}=$                             & $ M_8+M_9+ M_4 M_7$                                                                                                      \\
			$\frac{d \bm\theta(\lambda,\zeta)}{d B}=$                 & $\frac{d t_\gamma^{-1} (\psi)}{d \psi}\left(T_{B0}+T_{B1}+\right.$                                                       & $\frac{d \bm\theta(\lambda,\zeta)}{d \bm{d}}=$ & $\frac{d t_\gamma^{-1} (\psi)}{d \psi}\left(T_{d 0}+T_{d 1}+\right.$                                           \\
			& $\left.T_{B2}+T_{B3}\right)$                                                                                             &                                             & $\left.T_{d 2}+T_{d 3}\right)$                                                                                 \\ \hline
			\multicolumn{2}{l}{Computing $\frac{d \bm\theta(\bm{\varepsilon},\bm\lambda)}{d \bm{\alpha_\psi}}$}                                                                                                & \multicolumn{2}{l}{Computing $\nabla_\theta \log q_\lambda(\bm\theta)$}                                                                                                \\ \hline
			$M_1=$                                                 & $\sqrt{1-\bm{\tilde{\delta}}_\psi^\top\Sigma_\psi^{-1}\bm{\tilde{\delta}}_\psi}$                                            & $T_{q1} =$                                  & $(t_{\gamma_1}''(\theta_1)/t_{\gamma_1}'(\theta_1),\dots,t_{\gamma_m}''(\theta_m)/t_{\gamma_m}'(\theta_m))^\top$            \\
			$M_2=$                                                 & $ -M_1^{-1}\bm{\tilde{\delta}}_\psi^\top\Sigma_\psi^{-1}$                                                                   & $M_1=$                                      & $\text{diag}\left( t_{\gamma_1}'(\theta_1),\dots,t_{\gamma_m}'(\theta_m)\right)$                                         \\
			$M_3=$                                                 & $\varepsilon_0 \text{vec}(I_m)\otimes M_2$                                                                                 & $T_{q2} =$                                  & $-M_1^\top\Sigma_\psi^{-1}(\bm\psi-\bm\mu_\psi)$                                                                            \\
			$M_4=$                                                 & $\bm{\tilde{\delta}}_\psi^\top\otimes I_m$                                                                                  & $M_4=$                                      & $\bm\alpha_\psi^\top S_{\psi}^{-1/2}(\bm\psi-\bm\mu_\psi)$                                                                   \\
			$M_5=$                                                 & $  M_4M_3 +\varepsilon_0 M_1 I_m$                                                                                          & $T_{q3}=$                                   & $M_1^\top S_\psi^{-1/2} \bm\alpha_\psi \frac{\phi_1(M_4)}{\Phi_1(M_4)}$                                                     \\
			$M_6=$                                                 & $\bm{\xi}^\top\Sigma_\psi^{-1}\bm{\tilde{\delta}}_\psi I_m+\bm{\xi}^\top\Sigma_\psi^{-1}\otimes\bm{\tilde{\delta}}_\psi$       & $\nabla_\theta \log q_\lambda(\bm\theta)=$  & $T_{q1}+T_{q2}+T_{q3}$                                                                                                   \\
			$M_7=$                                                 & $ |r|I_m-M_6+ M_5$                                                                                                       &                                             &                                                                                                                          \\
			$M_8=$                                                 & $\frac{d t_\gamma^{-1}(\psi)}{d\psi} M_7S_\psi^{1/2}$                                                                    &                                             &                                                                                                                          \\
			$M_9=$                                                 & $1+\bm{\alpha_\psi}^\top\Omega_\psi\bm{\alpha_\psi}$                                                                        &                                             &                                                                                                                          \\
			$M_{10}=$                                              & $M_9^{-3/2}\left(M_9\Omega_\psi-\Omega_\psi\bm{\alpha_\psi}\bm{\alpha_\psi}^\top\Omega_\psi\right) $                        &                                             &                                                                                                                          \\
			$\frac{d \theta(\lambda,\zeta)}{d \bm{\alpha_\psi}} =$ & $M_8M_{10}$                                                                                                              &                                             &                                                                                                                          \\ \hline\hline
		\end{tabular}
	}
	\caption{Closed form expressions for four derivatives in Appendix~\ref{AppB}. 
		These are used to compute the gradient of the lower bound efficiently when
		using the reparameterization trick and a skew-normal copula approximation with a factor covariance structure. They are expressed recursively (with the terms evaluated from top to bottom for each derivative)
		and derived in the Online Appendix. In the table we denote $\bm\xi = \left(B\bm z+\bm{d}\circ\bm\epsilon\right)$, and $P$ is a matrix of zeros and ones such that $\frac{d \bm\theta(\bm{\varepsilon},\bm\lambda)}{d \bm{d}}=\frac{d \bm\theta(\bm{\varepsilon},\bm\lambda)}{d D}P$. MATLAB routines
	to evaluate the expressions are available in the Supplementary Materials.}
	\label{Tab:SkewNormDerivatives}
\end{table}

%

\newpage
\baselineskip=20pt
\noindent
\setcounter{page}{1}
{\bf \Large{Online Appendix for `High-dimensional Copula Variational Approximation through Transformation'
}}

\vspace{10pt}

\setcounter{figure}{0}
\setcounter{table}{0}
\renewcommand{\thetable}{A\arabic{table}}
\noindent
This Online Appendix has two parts:
\begin{itemize}
	\item[] {\bf Part~A}: Specifies the pair-copula used for the D-vine in Section 3.2.
	\item[] {\bf Part~B}: Derivation of four derivatives used in Appendix~B for applying the reparameterization trick to the skew-normal copula approximation. 
\end{itemize}
\newpage

\noindent {\bf \large{Part~A: Pair-copula Specification}}\\
\noindent 
Here, we specify the form of the pair-copula used to define the D-vine in Section 3.2.
\cite{LoaizaSmithManee2017} show that $c^{DV}$ is able to capture persistence
in the variance if one or more pair-copula $c_k$ allows for concentration of the probability mass in the four quadrants of the unit square. To do so they suggest
using the following mixture of rotated copulas for each of the pair-copulas (where we drop the subscript $k$ throughout for ease of presentation):
\begin{equation*}
c^{MIX}(u,v;\etavec) = w_1c^a(u,v;\etavec^a)+(1-w_1)c^b(1-u,v;\etavec^b)
\,,\;0\leq w_1 \leq1\,.
\end{equation*}
Here, the pair-copula parameter vector is $\etavec=\{\etavec^a,\etavec^b,w_1\}$, 
$w_1$ is a weight, and $c^a,c^b$ are two parametric bivariate copula densities with non-negative Kendall's tau and parameters $\etavec^a$ and $\etavec^b$ respectively. 
In the empirical work, for the mixture components $c^a$ and $c^b$ we employ the 
`convex Gumbel' defined as follows.
Let $c^G(u,v;\tau)$ be the density of a Gumbel copula parameterized (uniquely)
in terms of its Kendall tau value $0\leq \tau < 0.99$. (Note that we bound
$\tau$ away from 1 to enhance numerical stability of the D-vine copula.) 
Then the convex Gumbel has a density $c^{cG}$ equal to the convex
combination of that of the Gumbel and its rotation 180 degrees (ie. the
survival copula), so that
\[
c^{cG}(u,v;\tau,w_2)=w_2 c^G(u,v;\tau) + (1-w_2)c^G(1-u,1-v;\tau)\,,
\]
with $0\leq w_2 \leq 1$.
When employed for $c^a$ and $c^b$ it gives a
five parameter bivariate copula with $\etavec^a=(w_2^a,\tau^a)$, 
$\etavec^b=(w_2^b,\tau^b)$,
and a density $c^{MIX}$ that is equal 
to a mixture of all four 90 degree rotations of the Gumbel copula. 
We use independent uniform priors on the elements of $\etavec$ 
in our empirical work. 
\pagebreak

\noindent {\bf \large{Part~B: Variational approximation with skew-normal copula}}\\
\noindent
As shown in Section 4 and Appendix B, employing the skew-normal distribution for $\bm\psi=(\psi_1,\dots, \psi_m)^\top$, which we denote here as $\bm\psi\sim SN_m(\bm\mu_\psi,\Sigma_\psi,\bm\delta_\psi)$,  yields the following approximating density for $\bm{\theta}$
\begin{align*}
q_\lambda(\bm\theta) & = 2\phi_m(\bm\psi;\bm\mu_\psi,\Sigma_\psi)\Phi_1(\bm\alpha_\psi^\top S_\psi^{-1/2}\left(\bm\psi-\bm\mu_\psi)\right)\prod_{i=1}^m t_{\gamma_i}'(\theta_i)\,,
\end{align*}
The complete vector of variational parameters of this approximation,
$\bm\lambda=(\bm\mu_\psi^\top,\text{vech}(B)^\top,\bm d^\top,\bm\alpha_\psi^\top,\bm\gamma^\top)^\top$,  is obtained by optimizing the lower bound ${\cal L}(\bm\lambda)$ using SGA methods. As pointed out in Section 2, we obtain unbiased estimates of the gradient of ${\cal L}(\bm\lambda)$ by using the re-parametrization trick (Kingma and Welling, 2014; Rezende et al., 2014), in particular, we use the modification due to Roeder et al. (2017). To do this, we require the generative representation $\bm\theta=\bm\theta(\bm\varepsilon,\bm\lambda)=h\left(\bm\varepsilon,\bm\lambda\right)$, where $\bm\varepsilon$ is a vector of standardised random variables that have density  $f_{\bm{\varepsilon}}$ not depending on $\bm\lambda$. From this generative representation, we can then write the lower bound gradient
\begin{align}
\nabla_\lambda \calL(\lambdavec) =& E_{f_\varepsilon}\left[\nabla_\lambda \left(\log g(h(\varepsilonvec,\lambdavec))-\log q_\lambda(h(\varepsilonvec,\lambdavec))\right)\right]\nonumber\\
=&E_{f_\varepsilon}\left[\left\{\frac{d  h(\bm\varepsilon,\bm\lambda)}{d\bm\lambda}\right\}^T\left(\nabla_\theta \log g(h(\bm\varepsilon,\bm\lambda))-\nabla_\theta \log q_\lambda(h(\bm\varepsilon,\bm\lambda))\right)\right]\,,\label{roedergrad}
\end{align}
where unbiased estimates of $\nabla_\lambda {\cal L}(\bm\lambda)$ are obtained by drawing one or more Monte Carlo samples from $f_\varepsilon$ 
to approximate the expectation (Roeder et al., 2017). To derive the required generative relationship, $\bm\theta=\bm\theta\left(\bm\varepsilon,\bm\lambda\right)$, first, note that if $\bm\psi\sim SN_m(\bm\mu_\psi,\Sigma_\psi,\bm\delta_\psi)$, then we can think of $\bm\psi$ as arising from the following
generative model:
\begin{align}
r & \sim N(0,1)  \nonumber \\
\bm\psi|r &  \sim N(\bm\mu_\psi+\bm{\tilde{\delta}}_\psi |r|,\Sigma_\psi-\bm{\tilde{\delta}}_\psi \bm{\tilde{\delta}}_\psi^\top), \label{condsim}
\end{align}
where $\bm{\tilde{\delta}}_\psi=S_\psi^{1/2} \bm{\delta}_\psi$.  Note that if the conditional mean in (\ref{condsim}) were $\bm\mu_\psi+\bm{\tilde{\delta}}_\psi r$ rather than
$\bm\mu_\psi+\bm{\tilde{\delta}}_\psi |r|$ then the generative model above corresponds to $(r,\bm\psi)$ being jointly normal, 
$$N\left(\left[\begin{array}{cc} 0 \\ \bm\mu_\psi \end{array}\right],\left[\begin{array}{cc} 1 & \bm{\tilde{\delta}}_\psi^\top \\ \bm{\tilde{\delta}}_\psi & \Sigma_\psi \end{array}\right]\right).$$
The generative model for the skew normal can be regarded as conditioning $r\sim N(0,1)$ on $r>0$ and then generating from the conditional for $\bm\psi|r$ arising in the joint normal
distribution above.
The generative step (\ref{condsim}) can be written as 
\begin{align}
\bm\psi & =\bm\mu_\psi+\bm{\tilde{\delta}}_\psi|r| + (I-\bm{\tilde{\delta}}_\psi \bm{\tilde{\delta}}_\psi^\top \Sigma_\psi^{-1})\bm\xi +\sqrt{1-\bm{\tilde{\delta}}_\psi^\top \Sigma_\psi^{-1}\bm{\tilde{\delta}}_\psi}\bm{\tilde{\delta}}_\psi \varepsilon_0 \label{gencondsim}
\end{align}
where $\varepsilon_0\sim N(0,1)$ and $\bm\xi\sim N(\bm 0,\Sigma_\psi)$. Equation (\ref{gencondsim}) writes the conditional simulation step in (\ref{condsim}) in terms of 
a draw from the unconditional distribution $N(\bm 0,\Sigma_\psi)$ which allows us to make use of whatever structure is assumed for $\Sigma_\psi$ in applying the 
reparametrization trick.  To see that (\ref{gencondsim}) implements (\ref{condsim}) note that with $r$ fixed (i.e. conditional on $r$) we have
$\text{E}(\bm\psi)=\bm\mu_\psi+\bm{\tilde{\delta}}_\psi |r|$ and 
\begin{align*}
\text{Cov}(\psi) & = (I-\bm{\tilde{\delta}}_\psi \bm{\tilde{\delta}}_\psi \Sigma_\psi^{-1})\Sigma_\psi (I-\bm{\tilde{\delta}}_\psi \bm{\tilde{\delta}}_\psi^\top\Sigma_\psi^{-1})^\top+(1-\bm{\tilde{\delta}}_\psi^\top\Sigma_\psi^{-1}\bm{\tilde{\delta}}_\psi)\bm{\tilde{\delta}}_\psi \bm{\tilde{\delta}}_\psi^\top \\
& = \Sigma_\psi-\bm{\tilde{\delta}}_\psi \bm{\tilde{\delta}}_\psi^\top-\bm{\tilde{\delta}}_\psi \bm{\tilde{\delta}}_\psi^\top+\bm{\tilde{\delta}}_\psi \bm{\tilde{\delta}}_\psi^\top \Sigma_\psi^{-1} \bm{\tilde{\delta}}_\psi \bm{\tilde{\delta}}_\psi^\top
+\bm{\tilde{\delta}}_\psi \bm{\tilde{\delta}}_\psi^\top-\bm{\tilde{\delta}}_\psi^\top\Sigma_\psi^{-1}\bm{\tilde{\delta}}_\psi \bm{\tilde{\delta}}_\psi \bm{\tilde{\delta}}_\psi^\top \\
& = \Sigma_\psi-\bm{\tilde{\delta}}_\psi\bm{\tilde{\delta}}_\psi^\top
\end{align*}
upon observing that $\bm{\tilde{\delta}}_\psi^\top\Sigma_\psi^{-1}\bm{\tilde{\delta}}_\psi$ is a scalar.  
In the case of our factor parametrization where $\Sigma_\psi=BB^\top+D^2$, we can represent the draw $\bm\xi\sim N(\bm 0,\Sigma_\psi)$ as 
$$\bm\xi=B\bm z+\bm d\circ\bm\epsilon$$
where $\bm z=(z_1,\dots, z_p)^\top\sim N(\bm 0,I_p)$, $\bm\epsilon\sim N(0,I_m)$, $\bm z$ and $\bm\epsilon$ are independent, and $\circ$ denotes element by element (Hadamard) product
of two vectors.  So letting 
$\bm\varepsilon=(u,\bm z,\bm\epsilon,\varepsilon_0)\sim N(\bm 0,I_{m+p+2})$, we represent $q_\lambda(\bm\theta)$ as
\begin{align}
\bm\theta & = \bm\theta(\bm\varepsilon,\bm\lambda)\nonumber \\
& = t_\gamma^{-1}\left( \bm\mu_\psi+\bm{\tilde{\delta}}_\psi |r|+(I-\bm{\tilde{\delta}}_\psi \bm{\tilde{\delta}}_\psi^\top \Sigma_\psi^{-1})(B\bm z+\bm d\circ\bm\epsilon)+\sqrt{1-\bm{\tilde{\delta}}_\psi^\top \Sigma_\psi^{-1}\bm{\tilde{\delta}}_\psi}\bm{\tilde{\delta}}_\psi \varepsilon_0\right). \label{repartrick}
\end{align}
Finally, Equation (\ref{repartrick}) is the generative representation $\bm{\theta} = h\left(\bm\varepsilon,\bm\lambda\right)$ needed to derive closed-form expressions for the lower bound gradient in Equation \ref{roedergrad}.  
As shown in Appendix B, to evaluate (\ref{roedergrad}) it suffices to write down expressions for
$$\nabla_\theta \log q_\lambda(\bm\theta)\,,\;\;\frac{d \bm\theta(\bm\varepsilon,\bm\lambda)}{d \bm\mu_\psi}\,,\;\;\;\frac{d \bm\theta(\bm\varepsilon,\bm\lambda)}{dB}\,,\;\;\;\frac{d \bm\theta(\bm\varepsilon,\bm\lambda)}{d \bm d}\,,\;\;\;\frac{d \bm\theta(\bm\varepsilon,\bm\lambda)}{d \bm\gamma}\,,\;\ \text{and}\;\;\frac{d \bm\theta(\bm\varepsilon,\bm\lambda)}{d \bm\alpha_\psi}.$$
Notice that the term  $\nabla_\theta \log g(\bm\theta)=\nabla_\theta \log p(\bm\theta)p(y|\bm\theta)$ is model specific and needs to be considered on a case by case basis. The remainder of this Online Appendix is concerned with the derivation of close-form formulas for the expressions above. To this purpose, we will interchangeably use the symbol $\bm\xi$ to refer to  $\left(B\bm z+\bm d\circ\bm\epsilon\right)$.

Before deriving analytical expression to these gradient components, it is helpful at this point to establish some notation used in the derivations below.  For a $d$-dimensional vector valued function $g(\bm x)$ of an $n$-dimensional
argument $\bm x$, $\frac{d g}{d \bm x}$ is the $d\times n$ matrix with element $(i,j)$ $\frac{\partial g_i}{\partial x_j}$.  This means for a scalar $g(\bm x)$, $\frac{d g}{d \bm x}$ is
a row vector.  We write $\nabla_x g(\bm x)=\frac{d g}{d \bm x}^\top$.  When the function $g(\bm x)$ or the argument $\bm x$ are matrix valued, then $\frac{d g}{d \bm x}$ is taken to 
mean $\frac{d \text{vec}(g(\bm x))}{d \text{vec}(\bm x)}$, where $\text{vec}(A)$ denotes the vectorization of a matrix $A$ obtained by stacking its columns one
underneath another.  If $g(x)$ and $h(x)$ are matrix valued functions, say $g(x)$ takes values which are $d\times r$ and $h(x)$ takes values which are $r\times n$, 
then a matrix valued product rule is
\begin{align*}
\frac{d g(x)h(x)}{dx} & = (h(x)^\top\otimes I_d)\frac{d g(x)}{d x}+(I_n\otimes g(x))\frac{ d h(x)}{d x}
\end{align*}
where $\otimes$ denotes the Kronecker product and $I_a$ denotes the $a\times a$ identity matrix for a positive integer $a$.  
Some other useful results used repeatedly throughout the derivations below are
$$\text{vec}(ABC)=(C^\top\otimes A)\text{vec}(B),$$
for conformably dimensioned matrices $A$, $B$ and $C$
and 
\begin{align*}
\frac{d A^{-1}}{d A} & = -(A^{-T}\otimes A^{-1}).
\end{align*}
We also write $K_{m,n}$ for the commutation matrix (see, for example, Magnus and Neudecker, 1999).
\ \\
{\large\noindent {\bf Computing $\nabla_\theta \log q_\lambda(\bm\theta)$}}\\
\ \\
Noting that
\begin{align*}
q_\lambda(\bm\theta) & = \left\{\prod_{i=1}^m t_{\gamma_i}'(\theta_i)\right\} \times 2\phi_m(t_\gamma(\bm\theta);\bm\mu_\psi,\Sigma_\psi)\Phi_1(\bm\alpha_\psi^\top S_\psi^{-1/2}(t_\gamma(\bm\theta)-\bm\mu_\psi))
\end{align*}
we have
\begin{align*}
\log q_\lambda(\bm\theta) & =\sum_{i=1}^m \log t_{\gamma_i}'(\theta_i)+\log 2 + \log \phi_m(t_\gamma(\bm\theta);\bm\mu_\psi,\Sigma_\psi)+\log \Phi_1(\bm\alpha_\psi^\top S_\psi^{-1/2} (t_\gamma(\bm\theta)-\bm\mu_\psi))
\end{align*}
and hence
\begin{align*}
\nabla_\theta \log q_\lambda(\bm\theta) & = T_{q1}+T_{q2}+T_{q3}
\end{align*}
where
\begin{align*}
T_{q1} & = \sum_{i=1}^m \nabla_\theta \log t_{\gamma i}'(\theta_i) \\
& = (t_{\gamma_1}''(\theta_1)/t_{\gamma_1}'(\theta_1),\dots,t_{\gamma_m}''(\theta_m)/t_{\gamma_m}'(\theta_m))^\top,
\end{align*}
\begin{align*}
T_{q2} & = \nabla_\theta \log \phi_m(t_\gamma(\bm\theta);\bm\mu_\psi,\Sigma_\psi) \\
& = -\left\{\frac{d t_\gamma(\bm\theta)}{d\bm\theta}\right\}^\top\Sigma_\psi^{-1}(t_\gamma(\bm\theta)-\bm\mu_\psi),
\end{align*}
and
\begin{align*}
T_{q3} & = \nabla_\theta \log \Phi_1(\bm\alpha_\psi^\top S_\psi^{-1/2} (t_\gamma(\bm\theta)-\bm\mu_\psi)) \\
& = \left\{ \frac{d t_\gamma(\bm\theta)}{d\bm\theta}\right\}^\top S_\psi^{-1/2} \bm\alpha_\psi \frac{\phi_1(\bm\alpha_\psi^\top S_\psi^{-1/2}(t_\gamma(\bm\theta)-\bm\mu_\psi))}{\Phi_1(\bm\alpha_\psi^\top S_{\psi}^{-1/2}(t_\gamma(\bm\theta)-\bm\mu_\psi))}.
\end{align*}\\
\ \\
\noindent {\large{\bf Computing $\frac{d \bm\theta(\bm\varepsilon,\bm\lambda)}{d \bm\mu_\psi}$}}\\
\ \\
Writing
\begin{align*}
\bm\psi & = \bm\mu_\psi+\bm{\tilde{\delta}}_\psi |r|+(I-\bm{\tilde{\delta}}_\psi\bm{\tilde{\delta}}_\psi^\top\Sigma_\psi^{-1})\bm\xi+
\bm{\tilde{\delta}}_\psi \sqrt{1-\bm{\tilde{\delta}}_\psi^\top\Sigma_\psi^{-1}\bm{\tilde{\delta}}_\psi}\varepsilon_0,
\end{align*}
we have
\begin{align*}
\frac{d \bm\theta(\bm\varepsilon,\bm\lambda)}{d \bm\mu_\psi} & = \frac{d t_\gamma^{-1} (\bm\psi)}{d \bm\psi} \frac{d \bm\psi}{d\bm\mu_\psi}=\frac{d t_\gamma^{-1} (\bm\psi)}{d \bm\psi}.
\end{align*}\\
\ \\
\noindent {\large{\bf Computing }$\frac{d \bm\theta(\bm\varepsilon,\bm\lambda)}{d B}$}\\
\ \\
 This derivative can be written as
\begin{align*}
\frac{d \bm\theta(\bm\varepsilon,\bm\lambda)}{d B} & = \frac{d t_\gamma^{-1} (\bm\psi)}{d \bm\psi}\times \left\{T_{B0}+T_{B1}+T_{B2}+T_{B3}\right\}
\end{align*}
where

\begin{align*}
T_{B0} & = |r|\frac{d \bm{\tilde{\delta}}_\psi}{d B},
\end{align*}
where because  $\bm{\tilde{\delta}}_\psi=S_\psi^{1/2}\bm\delta_\psi=(\bm\delta_\psi^\top\otimes I_m)\text{vec}(S_\psi^{1/2})$, then
\begin{align*}
\frac{d \bm{\tilde{\delta}}_\psi}{d B} & = \frac{d \bm{\tilde{\delta}}_\psi}{d S_\psi^{1/2}}\frac{d S_\psi^{1/2}}{d S_\psi} \frac{d S_\psi}{d \Sigma_\psi}\frac{d \Sigma_\psi}{d B} \nonumber                                    
\end{align*}
By noticing that $\frac{d \Sigma_\psi}{d B}=\left(I_{m^2}+K_{m,m}\right)\left(B\otimes I_m\right)$, we can then compute $\frac{d \bm{\tilde{\delta}}_\psi}{d B}$ as
\begin{align}
\frac{d \bm{\tilde{\delta}}_\psi}{d B} & = \left(\bm\delta_\psi^\top \otimes I_m\right)\text{diag}\left(\text{vec}\left(\frac{1}{2}S_\psi^{-1/2}\right)\right)\text{diag}\left(\text{vec}\left(I_m\right)\right)\left(\left(I_{m^2}+K_{m,m}\right)\left(B\otimes I_m\right)\right),\nonumber \\
& = \text{diag}(\delta_\psi)S_\psi^{-1/2}[\text{diag}(B_{.1}),\dots \text{diag}(B_{.p})]\label{deriv1}
\end{align}
where for a vector $\bm a$, the function $\text{diag}(\bm a)$ is the diagonal matrix with diagonal entries $\bm a$. The columns of $B$ are denoted by $B_{.1},\dots, B_{.p}$ and 
in the expression $S_\psi^{-1/2}$ the power of the matrix is taken component wise. The next term is
\begin{align*}
T_{B1} & = \frac{d \bm\xi}{d B}= \bm z^\top\otimes I_m,
\end{align*}
which follows from noting that $\bm\xi=\left(\bm z^\top\otimes I_m\right)\text{vec}(B)+\bm d\circ\bm\epsilon$. The next term is $T_{B2}$
\begin{align*}
T_{B2} & =\bm{-}\frac{d \bm{\tilde{\delta}}_\psi \bm{\tilde{\delta}}_\psi^\top \Sigma_\psi^{-1}\bm\xi}{d B},
\end{align*}
The terms $T_{B2}$ can be computed as follows.
\begin{align}
T_{B2} & = \left(\bm\xi^\top\Sigma_\psi^{-1}\otimes I_m\right)\frac{d\bm{\tilde{\delta}}_\psi\bm{\tilde{\delta}}_\psi^\top}{d B}
+ \bm{\tilde{\delta}}_\psi \bm{\tilde{\delta}}_\psi^\top\frac{d\Sigma_\psi^{-1}\bm\xi}{d B} \label{TB2}
\end{align}
where 
\begin{align*}
\frac{d \bm{\tilde{\delta}}_\psi \bm{\tilde{\delta}}_\psi ^\top}{d B} & = \frac{d \bm{\tilde{\delta}}_\psi \bm{\tilde{\delta}}_\psi ^\top}{d \bm{\tilde{\delta}}_\psi} \frac{d \bm{\tilde{\delta}}_\psi}{d B} = (\bm{\tilde{\delta}}_\psi\otimes I_m+I_m\otimes \bm{\tilde{\delta}}_\psi)\frac{d \bm{\tilde{\delta}}_\psi}{d B}
\end{align*}
\begin{align}\label{Eq:complex1}
\frac{d \Sigma_\psi^{-1}\bm\xi}{d B} & = \left(\bm\xi^\top\otimes I_m\right) \frac{d\Sigma_\psi^{-1}}{d B}
+ \Sigma_\psi^{-1}\frac{d\bm\xi}{d B} 
\end{align}
and
\begin{align}
\frac{d \Sigma_\psi^{-1}}{d B} = -(\Sigma_\psi^{-1}\otimes \Sigma_\psi^{-1})\frac{d \Sigma_\psi}{d B}=-(\Sigma_\psi^{-1}\otimes \Sigma_\psi^{-1})\left(I_{m^2}+K_{m,m}\right)\left(B\otimes I_m\right).  \label{deriv2}
\end{align}

which can all be computed as $\frac{d\bm\xi}{d B}$, $\frac{d \bm{\tilde{\delta}}_\psi}{d B}$ and $\frac{d \Sigma_\psi}{d B}$ have been previously provided. 
The first term in $T_{B2}$ can be computed more efficiently by noticing that
\begin{align}
\left(\bm\xi^\top\Sigma_\psi^{-1}\otimes I_m\right)\frac{d\bm{\tilde{\delta}}_\psi\bm{\tilde{\delta}}_\psi^\top}{d B} =&\left(\bm\xi^\top\Sigma_\psi^{-1}\otimes I_m\right)(\bm{\tilde{\delta}}_\psi\otimes I_m+I_m\otimes \bm{\tilde{\delta}}_\psi)\frac{d \bm{\tilde{\delta}}_\psi}{d B}\nonumber\\
=& \left(\bm\xi^\top\Sigma_\psi^{-1}\bm{\tilde{\delta}}_\psi I_m+\bm\xi^\top\Sigma_\psi^{-1}\otimes\bm{\tilde{\delta}}_\psi\right)\frac{d \bm{\tilde{\delta}}_\psi}{d B}
\end{align}
So that 
\begin{align}
T_{B2} & = \left(\bm\xi^\top\Sigma_\psi^{-1}\bm{\tilde{\delta}}_\psi I_m+\bm\xi^\top\Sigma_\psi^{-1}\otimes\bm{\tilde{\delta}}_\psi\right)\frac{d \bm{\tilde{\delta}}_\psi}{d B}+ \bm{\tilde{\delta}}_\psi \bm{\tilde{\delta}}_\psi^\top\frac{d\Sigma_\psi^{-1}\bm\xi}{d B} \label{TB2}
\end{align}

The term $(\Sigma_\psi^{-1}\otimes \Sigma_\psi^{-1})$ can easily become computationally infeasible. To avoid using this term we compute the first term of Equation (\ref{Eq:complex1}) directly using a more simple expression. We use repeatedly the property of the commutation matrix that for $A_{m\times n}$ and $C_{r\times q}$ then 
$K_{r,m}(A\otimes C)=(C\otimes A)K_{q,n}$.  This means that 
\begin{align*}
(\Sigma_\psi^{-1}\otimes \Sigma_\psi^{-1})(I_{m^2}+K_{m,m}) = & (\Sigma_\psi^{-1}\otimes \Sigma_\psi^{-1}) +K_{m,m}(\Sigma_\psi^{-1}\otimes \Sigma_\psi^{-1}). \\
= & (I_{m^2}+K_{m,m})(\Sigma_\psi^{-1}\otimes \Sigma_\psi^{-1})
\end{align*}
Then using Kronecker product properties we can write
\begin{align*}
-(\Sigma_\psi^{-1}\otimes \Sigma_\psi^{-1})(I_{m^2}+K_{m,m})(B\otimes I_m) & = -(I_{m^2}+K_{m,m})(\Sigma_\psi^{-1}\otimes \Sigma_\psi^{-1})(B\otimes I_m) \\
& = -(I_{m^2}+K_{m,m})(\Sigma_\psi^{-1}B \otimes \Sigma_\psi^{-1}).
\end{align*}
From here we can then simplify the first term of Equation (\ref{Eq:complex1}) as
\begin{align*}
(\bm\xi^\top\otimes I_m)\frac{d \Sigma_\psi^{-1}}{d B} = & -(\bm\xi^\top\otimes I_m)(I_{m^2}+K_{m,m})(\Sigma_\psi^{-1}B \otimes \Sigma_\psi^{-1})                                                                                  \\
=                                                                         & -(\bm\xi^\top\Sigma_\psi^{-1}B \otimes \Sigma_\psi^{-1})                                -(\bm\xi^\top\Sigma_\psi^{-1}\otimes \Sigma_\psi^{-1}B)K_{m,p}  \\
=                                                                         & -(\bm\xi^\top\Sigma_\psi^{-1}B \otimes \Sigma_\psi^{-1})                                -K_{1,m}(\Sigma_\psi^{-1}B\otimes\bm\xi^\top\Sigma_\psi^{-1} )  \\
\end{align*}
Finally, for last term $T_{B3}$ we have that
\begin{align*}
T_{B3} = & \frac{d}{d B}\bm{\tilde{\delta}}_\psi \sqrt{1-\bm{\tilde{\delta}}_\psi^\top\Sigma_\psi^{-1} \bm{\tilde{\delta}}_\psi} \varepsilon_0 \\
= &(\bm{\tilde{\delta}}_\psi^\top\otimes I_m) \frac{d}{d B}\varepsilon_0\sqrt{1-\bm{\tilde{\delta}}_\psi^\top \Sigma_\psi^{-1}\bm{\tilde{\delta}}_\psi} I_m
+ \varepsilon_0 \sqrt{1-\bm{\tilde{\delta}}_\psi^\top \Sigma_\psi^{-1}\bm{\tilde{\delta}}_\psi} \frac{d\bm{\tilde{\delta}}_\psi}{d B},
\end{align*}
where $\frac{d \bm{\tilde{\delta}}_\psi}{d B}$ was computed previously and

For the first term in $T_{B3}$.  We have
\begin{align*}
& \left\{(\bm{\tilde{\delta}}_\psi^\top\otimes I_m)\right\} \frac{d}{d B}\varepsilon_0\sqrt{1-\bm{\tilde{\delta}}_\psi^\top \Sigma_\psi^{-1}\bm{\tilde{\delta}}_\psi} I_m \\
= & \left\{(\bm{\tilde{\delta}}_\psi^\top\otimes I_m)\right\} \text{vec}(I_m)\otimes \frac{d}{d B}\varepsilon_0 \sqrt{1-\tilde{\delta}^\top \Sigma_\psi^{-1} \bm{\tilde{\delta}}_\psi} \\
= & \bm{\tilde{\delta}}_\psi\otimes \frac{d}{d B}\varepsilon_0 \sqrt{1-\tilde{\delta}^\top \Sigma_\psi^{-1} \bm{\tilde{\delta}}_\psi} \\
= & -\varepsilon_0/2 (1-\bm{\tilde{\delta}}_\psi^\top\Sigma_\psi^{-1} \bm{\tilde{\delta}}_\psi )^{-1/2} \bm{\tilde{\delta}}_\psi\otimes \frac{d}{d B}\bm{\tilde{\delta}}_\psi^\top\Sigma_\psi^{-1} \bm{\tilde{\delta}}_\psi \\
= & -\varepsilon_0/2 (1-\bm{\tilde{\delta}}_\psi^\top\Sigma_\psi^{-1} \bm{\tilde{\delta}}_\psi )^{-1/2} \bm{\tilde{\delta}}_\psi\otimes \left\{\bm{\tilde{\delta}}_\psi^\top \frac{ d (\bm{\tilde{\delta}}_\psi^\top\Sigma_\psi^{-1})}{d B} + \bm{\tilde{\delta}}_\psi^\top\Sigma_\psi^{-1} \frac{ d \bm{\tilde{\delta}}_\psi}{d B}\right\}.
\end{align*}
In the above the calculations involving the second term in the sum can be done easily using our expression for $\frac{d \bm{\tilde{\delta}}_\psi}{d B}$ given earlier.  For the first term, we 
need to calculate
\begin{align*}
& -\varepsilon_0/2 (1-\bm{\tilde{\delta}}_\psi^\top\Sigma_\psi^{-1} \bm{\tilde{\delta}}_\psi )^{-1/2} \bm{\tilde{\delta}}_\psi\otimes \left\{\bm{\tilde{\delta}}_\psi^\top \frac{ d (\bm{\tilde{\delta}}_\psi^\top\Sigma_\psi^{-1})}{d B}\right\} \\
= & -\varepsilon_0/2 (1-\bm{\tilde{\delta}}_\psi^\top\Sigma_\psi^{-1} \bm{\tilde{\delta}}_\psi )^{-1/2} \bm{\tilde{\delta}}_\psi\otimes \left\{ \bm{\tilde{\delta}}_\psi^\top \left\{ \Sigma_\psi^{-1} \frac{d \bm{\tilde{\delta}}_\psi}{d B} +(I_m\otimes \bm{\tilde{\delta}}_\psi^\top)\frac{d \Sigma_\psi^{-1}}{d B} \right\}\right\}.
\end{align*}
Examining this last expression, the first term in the sum is easily computed using our expression for $\frac{d \bm{\tilde{\delta}}_\psi}{d B}$ given earlier, and it is only the second term 
that we need to worry about.  This second term is
\begin{align}
& -\varepsilon_0/2 (1-\bm{\tilde{\delta}}_\psi^\top\Sigma_\psi^{-1} \bm{\tilde{\delta}}_\psi )^{-1/2} \bm{\tilde{\delta}}_\psi\otimes \left\{ \bm{\tilde{\delta}}_\psi^\top \left\{ (I_m\otimes \bm{\tilde{\delta}}_\psi^\top)\frac{d \Sigma_\psi^{-1}}{d B} \right\}\right\} \label{Eq:complex2}\\
= & \varepsilon_0/2 (1-\bm{\tilde{\delta}}_\psi^\top\Sigma_\psi^{-1} \bm{\tilde{\delta}}_\psi )^{-1/2} \bm{\tilde{\delta}}_\psi\otimes \left\{ \bm{\tilde{\delta}}_\psi^\top  \left\{(I_m\otimes \bm{\tilde{\delta}}_\psi^\top)
(\Sigma_\psi^{-1}\otimes \Sigma_\psi^{-1})\right.\right. \nonumber\\ 
& \left.\left.\times (I_{m^2}+K_{m,m})(B\otimes I_m) \right\}\right\} \nonumber\\
= &  \varepsilon_0/2 (1-\bm{\tilde{\delta}}_\psi^\top\Sigma_\psi^{-1} \bm{\tilde{\delta}}_\psi )^{-1/2} \bm{\tilde{\delta}}_\psi\otimes \left\{ \bm{\tilde{\delta}}_\psi^\top \left\{ (I_m\otimes \bm{\tilde{\delta}}_\psi^\top)
(\Sigma_\psi^{-1}\otimes \Sigma_\psi^{-1})(B\otimes I_m) \right\}\right\} + \nonumber\\
& \varepsilon_0/2 (1-\bm{\tilde{\delta}}_\psi^\top\Sigma_\psi^{-1} \bm{\tilde{\delta}}_\psi )^{-1/2} \bm{\tilde{\delta}}_\psi\otimes \left\{ \bm{\tilde{\delta}}_\psi^\top \left\{(I_m\otimes \bm{\tilde{\delta}}_\psi^\top)
(\Sigma_\psi^{-1}\otimes \Sigma_\psi^{-1})(I\otimes B)K_{m,p} \right\} \right\} \nonumber\\
= & \varepsilon_0/2 (1-\bm{\tilde{\delta}}_\psi^\top \Sigma_\psi^{-1}\bm{\tilde{\delta}}_\psi)^{-1/2}\bm{\tilde{\delta}}_\psi \otimes \left\{\bm{\tilde{\delta}}_\psi^\top \left\{\Sigma_\psi^{-1}B\otimes \bm{\tilde{\delta}}_\psi^\top\Sigma_\psi^{-1}\right\} \right\}+ \nonumber\\
& \varepsilon_0/2 (1-\bm{\tilde{\delta}}_\psi^\top \Sigma_\psi^{-1}\bm{\tilde{\delta}}_\psi)^{-1/2}\bm{\tilde{\delta}}_\psi \otimes \left\{\bm{\tilde{\delta}}_\psi^\top \left\{\Sigma_\psi^{-1}\otimes \bm{\tilde{\delta}}_\psi^\top\Sigma_\psi^{-1} B\right\}K_{m,p}\right\},\nonumber
\end{align}
and this last expression is easily computable.
\ \\
\noindent {\large{\bf Computing $\frac{d \bm\theta(\lambda,\bm\varepsilon)}{d\bm d}$}}\\
\ \\
To compute $\frac{d \bm\theta(\lambda,\bm\varepsilon)}{d\bm  d}$ we notice first that
\begin{align*}
\frac{d \bm\theta(\lambda,\bm\varepsilon)}{d\bm d} & = \frac{d \bm\theta(\lambda,\bm\varepsilon)}{d D}P,
\end{align*}
where $P$ is the matrix of ones and zeros that extract columns $1,m+2,2m+3,\dots,m^2$ which correspond to the derivatives with respect to $\bm d$. Then, because of the symmetry in the way that $B$ and $D$ appear in 
$\Sigma_\psi=BB^\top+D^2=BB^\top+DD^\top$ the expression for
$\frac{d \bm\theta(\lambda,\bm\varepsilon)}{d D}$ is the same as that for
$\frac{d \bm\theta(\lambda,\bm\varepsilon)}{d B}$, except we need to replace 
all occurrences of $\frac{d \bm\xi}{dB}=\bm z^\top\otimes I_m$ by $\frac{d \bm\xi}{dD}=\bm\epsilon^\top\otimes I_m$, and replace $B$ with $D$, whenever $B$ appears outside the expression $\Sigma_\psi=BB^\top+D^2$.

Although the derivatives with respect to $B$ and $\bm d$ use equivalent formulas, some terms of $\frac{d \bm\theta(\lambda,\bm\varepsilon)}{d\bm d}$ need to be modified for computational efficiency.

The first expression we modify is 
$$\frac{d \Sigma_\psi^{-1}\bm\xi}{d \bm d} =( \bm\xi^\top\otimes I_m)\frac{d \Sigma_{\psi}^{-1}}{d D}P+ \Sigma_\psi^{-1}\frac{d\bm\xi}{d B}P$$
Following the same reasoning as in the previous section, the first term of this expression can be also written as 
\begin{align*}
( \bm\xi^\top\otimes I_m)\frac{d \Sigma_{\psi}^{-1}}{d D}P =  \left(-(\bm\xi^\top\Sigma_\psi^{-1}D \otimes \Sigma_\psi^{-1})-K_{1,m}(\Sigma_\psi^{-1}D\otimes\bm\xi^\top\Sigma_\psi^{-1} )\right)P
\end{align*}
Which means, we only need to compute columns $1,m+2,2m+3,\dots,m^2$ of $( \bm\xi^\top\otimes I_m)\frac{d \Sigma_{\psi}^{-1}}{d D}$. These columns can be individually computed by noticing that the jth column of 
$$-(\bm\xi^\top\Sigma_\psi^{-1}D \otimes \Sigma_\psi^{-1})                                -(\Sigma_\psi^{-1}D\otimes\bm\xi^\top\Sigma_\psi^{-1} )$$
is obtained as
$$-\left(\left[\bm\xi^\top\Sigma_\psi^{-1}D\right]_k\Sigma_{\psi,.k}^{-1}\right)-\left(\Sigma_\psi^{-1}D_{.k}\left[\bm\xi^\top\Sigma_\psi^{-1}\right]_k\right)$$
where $k =\frac{j+m}{1+m}$, $A_{.k}$ denotes the kth
column of any matrix $A$, $\left[\bm y\right]_k$ denotes the kth element of any vector $\bm y$ and $\Sigma_{\psi,.l}^{-1}$ denotes the
lth column of $\Sigma_{\psi}^{-1}$.

The second expression that we re-write is that in Equation~(\ref{Eq:complex2}). Specifically, we compute the value of $(I_m\otimes \bm{\tilde{\delta}}_\psi^\top)\frac{d \Sigma_\psi^{-1}}{d D}P$ directly, by noticing again that only the columns $1,m+2,2m+3,\dots,m^2$ of $(I_m\otimes \bm{\tilde{\delta}}_\psi^\top)\frac{d \Sigma_\psi^{-1}}{d D}$ are needed. These columns can be individually computed by noticing that the jth column of
$$(I_m\otimes \bm{\tilde{\delta}}_\psi^\top)\frac{d \Sigma_\psi^{-1}}{d D}=-\Sigma_\psi^{-1}D\otimes\bm{\tilde{\delta}}_\psi^\top\Sigma_\psi^{-1}-\bm{\tilde{\delta}}_\psi^\top\Sigma_\psi^{-1}D\otimes\Sigma_\psi^{-1}$$
is obtained as
$$-\left(\Sigma_\psi^{-1}D_{.k}\left[\bm{\tilde{\delta}}_\psi^\top\Sigma_\psi^{-1}\right]_k\right)-\left(\left[\bm{\tilde{\delta}}_\psi^\top\Sigma_\psi^{-1}D\right]_k\Sigma_{\psi,.k}^{-1}\right)$$\\
\ \\
\noindent {\large{\bf Computing $\frac{d \bm\theta(\lambda,\bm\varepsilon)}{d \bm\alpha_\psi}$}}\\
\ \\
To compute $\frac{d \bm\theta(\lambda,\bm\varepsilon)}{d \bm\alpha_\psi}$ notice that
\begin{align*}
\frac{d \bm\theta(\lambda,\bm\varepsilon)}{d \bm\alpha_\psi} & =\frac{d \bm\theta(\lambda,\bm\varepsilon)}{d \bm\delta_\psi}\frac{d \bm\delta_\psi}{d \bm\alpha_\psi},
\end{align*}
we know that $$\bm\delta_\psi=\frac{1}{\left(1+\bm\alpha_\psi^\top\Omega_\psi\bm\alpha_\psi\right)^{1/2}}\Omega_\psi\bm\alpha_\psi$$
therefore
\begin{align*}
\frac{d \bm\delta_\psi}{d \bm\alpha_\psi} & =\frac{1}{\left(1+\bm\alpha_\psi^\top\Omega_\psi\bm\alpha_\psi\right)^{3/2}}\left(\left(1+\bm\alpha_\psi^\top\Omega_\psi\bm\alpha_\psi\right)\Omega_\psi-\Omega_\psi\bm\alpha_\psi\bm\alpha_\psi^\top\Omega_\psi\right),
\end{align*}
and
\begin{align*}
\frac{d \bm\theta(\lambda,\bm\varepsilon)}{d \bm\delta_\psi} & = \frac{d t_\gamma^{-1}(\bm\psi)}{d\bm\psi} \frac{d \bm\psi}{d\bm{\tilde{\delta}}_\psi}\frac{d \bm{\tilde{\delta}}_\psi}{d \bm\delta_\psi},
\end{align*}
where 
\begin{align*}
\frac{d \bm{\tilde{\delta}}_\psi}{d \bm\delta_\psi} & = S_\psi^{1/2},
\end{align*}
\begin{align*}
\frac{d \bm\psi}{d \bm{\tilde{\delta}}_\psi} & = |r|I_m-\frac{d}{d\bm{\tilde{\delta}}_\psi} \left\{ \bm{\tilde{\delta}}_\psi \bm{\tilde{\delta}}_\psi^\top\Sigma_\psi^{-1}\bm\xi\right\} \\
& + \frac{d}{d\bm{\tilde{\delta}}_\psi} \left\{ \sqrt{1-\bm{\tilde{\delta}}_\psi^\top \Sigma_\psi^{-1}\bm{\tilde{\delta}}_\psi} \varepsilon_0 \bm{\tilde{\delta}}_\psi \right\},
\end{align*}
\begin{align*}
\frac{d}{d\bm{\tilde{\delta}}_\psi} \left\{ \bm{\tilde{\delta}}_\psi \bm{\tilde{\delta}}_\psi^\top\Sigma_\psi^{-1}\bm\xi\right\} & = \left\{\bm\xi^\top \Sigma_\psi^{-1}\otimes I_m\right\}\frac{d}{d \bm{\tilde{\delta}}_\psi} \bm{\tilde{\delta}}_\psi \bm{\tilde{\delta}}_\psi^\top \\
& = \left\{\bm\xi^\top\Sigma_\psi^{-1}\otimes I_m\right\}\left\{\bm{\tilde{\delta}}_\psi \otimes I_m+I_m\otimes \bm{\tilde{\delta}}_\psi\right\},\\
&= \bm\xi^\top\Sigma_\psi^{-1}\bm{\tilde{\delta}}_\psi I_m+\bm\xi^\top\Sigma_\psi^{-1} \otimes \bm{\tilde{\delta}}_\psi
\end{align*}
\begin{align*}
\frac{d}{d\bm{\tilde{\delta}}_\psi} \left\{ \sqrt{1-\bm{\tilde{\delta}}_\psi^\top \Sigma_\psi^{-1}\bm{\tilde{\delta}}_\psi} \varepsilon_0 \bm{\tilde{\delta}}_\psi \right\} & = 
\left\{\bm{\tilde{\delta}}_\psi^\top\otimes I_m\right\} \frac{d}{d\bm{\tilde{\delta}}_\psi} \varepsilon_0\sqrt{1-\bm{\tilde{\delta}}_\psi^\top\Sigma_\psi^{-1} \tilde{\delta}\psi} I_m \\
& +
\varepsilon_0\sqrt{1-\bm{\tilde{\delta}}_\psi^\top\Sigma_\psi^{-1}\bm{\tilde{\delta}}_\psi} I_m,
\end{align*}
\begin{align*}
\frac{d}{d\bm{\tilde{\delta}}_\psi} \varepsilon_0\sqrt{1-\bm{\tilde{\delta}}_\psi^\top \Sigma_\psi^{-1} \bm{\tilde{\delta}}_\psi} I_m & = \varepsilon_0 \text{vec}(I_m)\otimes \frac{d}{d\bm{\tilde{\delta}}_\psi}
\sqrt{1-\bm{\tilde{\delta}}_\psi^\top \Sigma_\psi^{-1} \bm{\tilde{\delta}}_\psi},
\end{align*}
\begin{align*}
\frac{d}{d\bm{\tilde{\delta}}_\psi} \sqrt{1-\bm{\tilde{\delta}}_\psi^\top\Sigma_\psi^{-1} \bm{\tilde{\delta}}_\psi} & = 
\frac{- \bm{\tilde{\delta}}_\psi^\top\Sigma_\psi^{-1}}{\sqrt{1-\bm{\tilde{\delta}}_\psi^\top\Sigma_\psi^{-1}\bm{\tilde{\delta}}_\psi}}.
\end{align*}\\
\ \\
\noindent {\large{\bf Computing $\frac{d \bm\theta(\lambda,\bm\varepsilon)}{d \bm\gamma}$}}\\
\ \\
Finally, 
\begin{align*}
& \frac{d \bm\theta(\lambda,\bm\varepsilon)}{d \bm\gamma}
\end{align*}
is an $m$ by $m$ diagonal matrix, with $j$th diagonal element
\begin{align*}
& \frac{d t_{\gamma_i}^{-1}(\psi_i)}{d \gamma_i}
\end{align*}
which just involves differentiating the Yeo-Johnson transformation with respect to its parameter.  
\ \\ 
\ \\
\textbf{Online Appendix References}\ \\
\ \\
Azzalini, A. and Capitanio, A. (2003). Distributions generated by perturbation of symmetry with
emphasis on a multivariate skew t-distribution. Journal of the Royal Statistical Society: Series B
(Statistical Methodology), 65(2):367{389}.\ \\
	\ \\
Kingma, D. P. and Welling, M. (2014). Auto-encoding variational Bayes. In Proceedings
of the 2nd International Conference on Learning Representations (ICLR) 2014. https:
	arxiv.org/abs/1312.6114.\\
	\ \\
Magnus, J. and Neudecker, H. (1999). Matrix Differential Calculus with Applications in
Statistics and Econometrics. Wiley Series in Probability and Statistics. Wiley.\\ 
	\ \\
Rezende, D. J., Mohamed, S., and Wierstra, D. (2014). Stochastic backpropagation and
approximate inference in deep generative models. In Xing, E. P. and Jebara, T., editors,
Proceedings of the 29th International Conference on Machine Learning, ICML 2014.
proceedings.mlr.press/v32/rezende14.pdf.\\
	\ \\
Roeder, G., Wu, Y., and Duvenaud, D. (2017). Sticking the landing: Simple, lower-variance gradient
estimators for variational inference. arXiv:1703.09194.

\end{document}